\documentclass[nofootinbib,aps,prd,10pt,superscriptaddress,showpacs,preprintnumbers,eqsecnum]{revtex4-1}

\usepackage{amsmath}
\usepackage{dsfont}
\usepackage{multirow}
\usepackage{graphicx}
\usepackage{mathrsfs}
\usepackage{amsfonts}
\usepackage{amssymb}
\usepackage[dvips]{color}

\begin{document}
\newcommand\be{\begin{equation}}
\newcommand\ee{\end{equation}}
\newcommand\bea{\begin{eqnarray}}
\newcommand\eea{\end{eqnarray}}
\newcommand\bseq{\begin{subequations}} %solo con amsmath
\newcommand\eseq{\end{subequations}}
\newcommand\bcas{\begin{cases}}
\newcommand\ecas{\end{cases}}
\newcommand{\p}{\partial}
\newcommand{\f}{\frac}

\title{Stochastization of BKL dynamics and Anisotropic Sky Patterns}

\author{Orchidea Maria Lecian}
\email{lecian@icra.it}
\affiliation{Sapienza University of Rome, Physics Department and ICRA-International Center for Relativistic Astrophysics,\\ Piazzale Aldo Moro, 5- 00185 Rome, Italy}

\today

\begin{abstract}
The dynamics of cosmological billiards in $4=3+1$ spacetime dimensions is analyzed; the different statistical maps are characterized within the stochastic limit, reached after a large number of iterations of the billiard maps.\\
New densities of invariant measures have been established, also for billiard systems which contain symmetry walls, according to the content of Weyl reflections in the maps, which account for the change of sign of the non-oscillating scale factors in the solution to the Einstein field equations in the asymptotic limit towards the cosmological singularity. The statistical equivalence between the big billiard and the small billiard, posed in [Phys. Rev. D83, 044038 (2011)], is here proven by means of these new definitions of probabilities for the small billiard following the symmetries defined in the analysis of the systems on the Upper Poincar\'e Half Plane.\\ 
Further new classes of BKL probabilities have also been defined especially for the one-variable map and for the two-variable map, for the early-time BKL dynamics, for a stochastizing BKL dynamics and for a completely stochastized dynamics, both for the big billiard and for the small billiard. The trajectories have been classified according to these new probabilties, and different specifications of probabilties comparing classes of initial conditions have been assigned for the stochastization of the dynamics.\\
As a result, is is possible to establish a definition of BKL probabilities for the unquotiented dynamics of the big billiard, where the different patterns of Weyl reflections are encoded. The statistical description of BKL probabilities for the occurrence of a given number of epochs in each era are therefore further characterized by the most probable number of Weyl reflections contained in such eras, which is inferred from the implications of the billiard maps on the UPHP.\\
These new constructions have been considered for the determination of the connection between the observed values of anisotropy by a stochastic limit of the BKL dynamics.
\end{abstract}


\maketitle
\section{Introduction\label{section1}} 
Cosmological billiards arise as the description of the features of space-time in the asymptotic limit towards the cosmological singularity under the BKL (Belinskii, Khalatnikov–Lifshitz) hypothesis, \cite{KB1969a},\cite{Khalatnikov:1969eg},\cite{BK1970},\cite{BLK1971},\cite{Lifshitz:1963ps},\cite{LLK}, for which spacetime points are spatially decoupled within this limit, and the Einstein field equations reduce to s system of ordinary differential equations with respect to time, as time derivatives dominate the dynamic The chaotic motion of a billiard ball in a billiard system, which follows the geodesic evolution of bounces with respect to the (in the limit) infinite potential walls which define the billiard table, is the asymptotic description of the Bianchi IX cosmological model \cite{misn}\cite{Misner:1969hg},\cite{chi1972},\cite{Misner:1994ge} to which the most general anisotropic and homogeneous models are schematized under the BKL paradigm, by the definition of the appropriate statistical maps \cite{Chernoff:1983zz},\cite{isnai83},\cite{isnai85}.\\
The original BKL picture concerns the case of pure gravity in $4=3+1$ space-time dimensions. When also the asymptotic limit of more general inhomogeneous models are dealt with, the appearance of the so-called symmetry walls defines a different (smaller) kind of billiard. For this, one usually refers to the big billiard and the small billiard within all these specifications. The BKL paradigm has proven extremely successful in the description of higher-dimensional systems arising from higher-dimensional unification theories, where new geometrical structures are present, and where a discussion of the physical interpretation of such solutions is based on the proper BKL limit, for which the usual $4=3+1$ dimensional description results as the suitable limit for those physical systems, where a geometry based on the suitable algebraic structures is hypothesized for the target space in which the solution of the Einstein field equations can be represented \cite{Damour:2001sa},\cite{Damour:2000hv},\cite{Damour:2002fz},\cite{Damour:2002cu},\cite{Damour:2002et},\cite{hps2009}. A precise characterization of the $4=3+1$ model descending from these structures has recently been achieved within the framework the the billiard description of the dynamics, for which several symmetry-quotienting mechanisms have been defined according to the geometrical features of the space where the billiards are represented, and according to the Hamiltonian description of the corresponding dynamical systems \cite{Damour:2010sz},\cite{lecianproc}, \cite{Lecian:2013cxa}.\\
\\
Billiard systems on the UPHP have been studied, from a statistical point of view, in a 'mathematical' characterization, by several authors for billiard on the UPHP \cite{Balazs:1986uj},\cite{Bogomolny:1992cj},\cite{bogomolny1}, \cite{venkov1982}, \cite{heller}, \cite{berry}, \cite{berri}, \cite{matrix1}, \cite{matrix2}. The specific nature of orbits of billiards on the UPHP for arithmetical groups has been investigated in \cite{Bogomolny:1992cj} \cite{arith1} \cite{arith2}. In these works, the group-theoretical \cite{terras} interpretation of the properties of the generators of the billiard map has been favored with respect to the physical interpretation of the billiard dynamics.\\
\\
The need of the classification of trajectories in between the comparison of the original BKL description and the discovery of new billiard structures DAMOURSPINDEL  has been stressed in \cite{bel2009}. An analysis of the statistical properties underlying the dynamics of cosmological billiard systems on the UPHP has been introduced in \cite{cornish}, and in \cite{levin2000} for the quantum version, and compared in \cite{lecian}. The examination of the statistical features of periodic irrationals is still an open project in modern number theories, as the extent of the validity of the Gauss-Kuzmin theorem for this case is still under investigation \cite{mnt}.\\
\\
The chaotic dynamics of the asymptotic Bianchi IX systems in $4=3+1$ dimensions have been classified according to their metric entropy and to their topological entropy in \cite{barrowpr}.\\
\\
The relevance of the classification of trajectories for cosmological billiards is that several of their statistical features are invariant under the billiard maps, and their validity holds at the classical level, at the quantum regime and at the semiclassical limit. From the quantum point of view, the features of the wavefunction of the universe has been shown to be possibly connected with the large-scale structure of the universe \cite{levin2000}, from an analysis of its features on the UPHP, while the well-behaviored-ness of these functions before the implementation of the hamiltonian constraint has been ensured in \cite{Kleinschmidt:2009cv}, \cite{mk11}. This enforces the comparison of quantum gravity effects motivated by the effects of unification theories \cite{amati},\cite{8} and by different formulations of General Relativity at the Planck scale \cite{ash2},\cite{ash1},\cite{am1}, and by modifications of quantum mechanics in the strong-gravitational filed limit \cite{maggiore},\cite{altro} with the present observations \cite{9a},\cite{9b},\cite{9c}.\\
\\
Different perspectives in the characterization of the cosmological singularity have borough a thorough set of results during the last decades.\\
The description of the cosmological singularity has also been accomplished within the conformal Hubble-normalized orthonormal frame variables, in the dynamical-system approach, where the physical characterization of spacetime is achieved by a state space associated to a state vector composed of the diagonal components of the traceless shear matrix, the Fermi rotation variables, and the spatial commutation functions (i.e. the connections) that describe the three-curvature associated to the conformal metric: a picture results, \cite{wain} \cite{uggla2003} \cite{uggla2013}, which is, to a precise extent, dual to that illustrated by the choice of Iwasawa variables, as far as the description of the kasner circle is concerned. The curvature scalar and the properties of the geodesics have been analyzed, within the implications of this framework, in \cite{ringstrom2000} and \cite{Ringstrom:2000mk}, \cite{Heinzle:2009eh}.\\
A different group-theoretical assumption has been made in \cite{Berger:2000hb}.\\
Numerical simulations of the behavior of a generic universe in cosmology have been presented in \cite{garf93}, \cite{berger1993} \cite{berger1994}, \cite{berger1997}.\\
The presence of spikes \cite{uggla2012} can find interesting group-theoretical explanations within the framework of solution-generating techniques within  the physical interpretations of the structure constants of the Bianchi classification \cite{wct}, \cite{uggla2013} and also as far as the Petrov classification is concerned \cite{Bini:2008qg}, \cite{Cherubini:2004yi}, while a physical characterization of the algebraic properties of physical cosmological billiards has been provided in \cite{Fleig:2011mu}, \cite{carb}, to which the discussion of \cite{berger1991} could apply. Cosmological billiards delimited by finite-height potential walls are obtained in the higher-dimensional models of \cite{fre1},\cite{fre2}.\\
From a quantum point of view, the consistency of the mathematical features of the wavefunction before the solution of the Hamiltonian constraint has been established in \cite{Kleinschmidt:2009cv}, \cite{mk11}, while the mathematical features of the WDW equations have been defined in \cite{primordial}, \cite{puzio}, \cite{monix}, and the interpretative problems rised by the corresponding wavefucntion have been set in \cite{gibb}, \cite{Graham:1990jd}, \cite{isham1} \cite{isham}. A characterization of the wavefunction with respect to the anisotropy of the cosmological model is found in \cite{hawking84}, \cite{amsterdamski85}, \cite{moss85}, \cite{furusawa86}, \cite{furusawa861}, \cite{berger}, while a characterization of the wavefucntion with respect to the boundary conditions can be found in \cite{csordas1991}, \cite{Kleinschmidt:2009hv}, \cite{Forte:2008jr}, \cite{Graham:1990jd}, \cite{graham1991},\cite{Kleinschmidt:2010bk}, and in \cite{Benini:2006xu} a wavefunction on a distorted domain is considered.\\
\\
The continuous billiard dynamics can be described by the analysis of the discrete Poincar\'e return map of the billiard ball on a suitable Poincar\'e surface of section, for which the BKL maps and the CB-LKSKS maps are obtained. From a different point of view, a similar connection to the interpretation of the solution of the Einstein equations as those for a billiard system from a different perspective are obtained in \cite{lecian}.\\
In the present work, the set of initial conditions which physically characterize to solution to the Einstein field equations for the asymptotic limit towards the cosmological singularity under the BKL paradigm are classified according to their implications about the sequence of Kasner eras which they describe: in particular, singular trajectories, periodic trajectories and infinite non-periodic trajectories are analyzed.\\
Even though singular trajectories and periodic trajectories constitute countable sets, a non-zero probability for these phenomena is obtained within the framework of the statistical maps which describe the evolution of the dynamics in cosmological billiards. More in particular, the density of the invariant measure of the one-variable maps is used as a probability mass function for the discrete probability distribution related to these countable sets \cite{kolmogorovprobability}. It is crucial to remark that the possibility to implement new definitions of invariant measures and probabilities for cosmological billiard has already been established in \cite{Damour:2010sz} for the case of the epoch maps.\\
\\
These characterization of the initial conditions can be considered as 'complementary' to the analysis of periodic trajectories and of singular trajectories envisaged by the definition of Farey maps. Differently from the definition of Farey maps, the classification approached in the present analysis does not make any  assumption on the statistical maps, and a natural definition for probabilities for these particular trajectories arises.\\
For the present analysis, the statistical properties of the billiard maps have been analyzed by the definition of new densities of measure invariant under the billiard maps, which are based on the number of Weyl reflections, which account for the change of sign of the derivative of the non-oscillating scale factor of the solution to the Einstein field equations. These densities of measures produce new definitions for probabilities. The newly-defined probabilities constitute the 'building blocks' for the definition of other kinds of probabilities, which encode the statistical meaning of the different billiard maps, and exhibit particular features in the stochastization of the dynamics.
\\
The paper has been organized as follows.\\
In Section \ref{section2}, the main features of cosmological billiards in $4=3+1$ spacetime dimensions have been recalled.\\
In Section \ref{section3}, orbits and sequences of eras have been classified. Particular attention has been paid to singular trajectories and to periodic trajectories, as they are defined by initial conditions consisitng of countable sets of discrete variables.\\
In Section \ref{section4}, densities of invariant measures for the billiard dynamics have been established. After having recalled the original definition of density of invariant measure, which is due to the analysis \cite{KB1969a}, \cite{Khalatnikov:1969eg}, \cite{BK1970}, \cite{BLK1971} \cite{Lifshitz:1963ps}, \cite{LLK}, \cite{isnai83}, \cite{isnai85} , for the big billiard, the definition for invariant densities of measure for the iterations of the small billiard map have been established, according to the analysis of the small billiard map on the UPHP.\\
In Section \ref{section5}, new probabilties for the big billiard have been defined, which are based on the physical interpretation of the versions of the BKL map and of the CB-LKSKS maps, i.e. the one-variable map and the two-variable map, according to the role of the 'storage' of all the information of the exact evolution of the big billiard in one statistical variable, which allows one to define the stationary limiting distributions for the statistical maps.\\
In Section \ref{section6}, the possible steps which define the stochastization of the dynamics of the big billiard have been outlined, by means of the investigation tools achieved in the previous Sections.\\
In Section \ref{section7}, the full stochastized regime for the BKL dynamics of the big billiard has been analyzed according to the statistical maps.\\
In Section \ref{section8}, the new BKL probabilities for the sequence of epochs and eras of the small billiard have been evaluated according to the new densities of invariant measure for the small billiard maps defined in Section \ref{section4}, and the new BKL probabilities for the different versions of the BKL map and of the CB-LSKSK map, which differ according to the role of the 'memory' of the small-billiard system about the 'past' evolution, have been also defined.\\
In Section \ref{section9}, the steps of the transition of the BKL dynamics of the small billiard towards the fully-stochastized limit have been depicted, and  comparisons with the same transformations of the dynamics of the big billiard have been outlined as far as the features qualifying the small billiard within this regimes are concerned.\\
In Section \ref{section10}, the fully-stochastized dynamics of the small billiard has been described, and the several differences which distinguish this model form the stochastic limit of the original BKL dynamics have been analyzed.\\
In Section \ref{section11}, a comparison of the big billiard map and of the small billiard maps has been accomplished. The statistical equivalence between the two models has been proven, thus providing the evidence requested in \cite{Damour:2010sz}, both for the early time BKL dynamics, and for the fully stochastized dynamics, and both for the exact BKL probabilities and for the limit of BKL probabilities for long eras. The prof of the equivalence has also been nicely complemented by the definition of a further quantity qualifying the original BKL probabilities for the pure gravitational case, i.e. the number of Weyl reflections in the corresponding stages of the small billiard system, as classified in \cite{Kleinschmidt:2009cv}  for the most general description of the dynamic of cosmological billiards up to $11=10+1$ spacetime dimensions, and explicitly specified for the case $4=3+1$ in \cite{Lecian:2013cxa}.\\
In Section \ref{section12}, singular trajectories for cosmological billiards have been characterized according to the analysis developed in the previous section, as far as the relevance of billiard trajectories is concerned for the comparison of the original BKL description and the description of the mew structures discovered in higher-dimensional unification theories, as outlined in \cite{bel2009}.\\
In Section \ref{section13}, the connection has been defined, between the detection of small anisotropy in the actual experimental evidence for the CMB and the age of the universe at which a suitable quasi-isotropization mechanism has to be applied for the obtention of the actual description of the present universe, through the estimation of the corresponding degree of stochasticity reached by the BKL dynamics of the very early universe, regardless to the physical regime (i.e. the quantum regime, its semiclassical limit or its classicized outcome) such a quasi-isotropization mechanism has come into action. This most general connection has been rendered possible by the independence of the interpretation of the BKL trajectories of the (possible) quantum features of the gravitational interaction below the Planck scale.\\
Brief concluding remarks end the paper in Section \ref{section14}.
%%%%%%%%%%%%%%%%%%%%%%%%%%%%%%%%%%%%%% 
\section{Basic statements\label{section2}}
According to the BKL paradigm, space-time points can be considered as spatially decoupled in the asymptotic limit close to the cosmological singularity: within this limit and under this assumption, the general cosmological solutions acquire, in the asymptotic limit, the same asymptotic limit of the Bianchi IX model.\\
Within this framework, the Einstein filed equations reduce to a system of ODE for the time variable, whose solution can be approximated to that of a succession of Kasner solutions.\\
These analysis have been first performed in $4=3+1$ space-time dimensions; recent results in higher-dimensional unification theories have used this paradigm also in the higher-dimensional case.\\
For a simple Kasner solution in $4=3+1$, the three Kasner exponents can be ordered such that the first two define the expanding (and oscillating) directions, while the third one accounts for the (non-oscillating) contracting direction.\\ 
In the case of the asymptotic limit of the Bianchi IX model, the unordered triple of the Kasner scale factors, which play the role of Kasner coefficients, encodes the symmetries of the metric tensor, i.e. the features of the space-time. Each new Kasner parametrization in terms of a different Kasner solution for the asymptotic Bianchi IX case corresponds to a different (re)-ordering of the triple of Kasner coefficients, which is the result of the action of a symmetry transformation. In particular, the three Kasner exponents, both in the case of the simple Kasner solution and in the case of the Bianchi IX solution, obey the constraints
\begin{equation}\label{eq1}
\sum_ip_i=\sum_ip_i^2=1,
\end{equation}
such that can be parameterized in terms of one (auxiliary) variable $u$, and
can be ordered such that
\begin{equation}\label{eq2} 
p_1(u)>p_2(u)>p_3(u).
\end{equation}
The logarithmic scale factors $\beta$ describe the solution of the Einstein equations as the chaotic motion of a point particle in a Minkoskian space endowed with Lorentz symmetries (where geodesics are straight lines) bouncing inside the surface of a unit hyperboloid. The elimination of one degree of freedom  by the solution of the Hamiltonian constraint corresponds, after the definition of the suitable variables, to the projection of the chaotic motion onto the surface of the unit hyperboloid, where a two-dimensional billiard system is obtained as one delimited by the intersection of the potential walls with the unit hyperboloid. Considering only gravitational walls is equivalent to considering the most general anisotropic solution, while considering also the presence of symmetry walls corresponds to the more general inhomogeneous case. (It is important to recall that, in the higher-dimensional cases, the domains of the billiard table is defined also by considering different contributions to the Einstein field equations).\\
The two-dimensional billiard motion on the unit hyperboloid can be visualized, after the suitable geometrical transformations, as one on the unit circle, or as one on a triangular domain on the UPHP, where geodesics are (generalized) half circles centered on the axises axes. The cosmological singularity corresponds to the boundary of the unit circle, and to the horizontal axes $u$ of the UPHP (coordinatized by the complex variable $z=u+iv$) plus one point at infinity. These features are illustrated in Figure \ref{figura1} and further explained in the next Subsections.\\
For a generic two-dimensional system, the phase space is four-dimensional. Nonetheless, by fixing a particular energy shell, at which $E=const$, for which the dynamics on the billiard table is described by geodesic evolution and elastic bounces on the billiard ball, and by considering (a suitable parametrization of) the Poincar\`e return map of the evolution of the billiard ball on a suitable surface of section, two degrees of freedom are eliminated, such that a two-dimensional reduced phase space is obtained, where all the information about this schematization of the dynamics is encoded. The reduced phase space is coordinatized by \cite{Damour:2010sz} the variables $u^+$ and $u^-$, which are the oriented endpoints of the geodesics followed by the trajectories of the billiard ball during the 'free-flight' evolution between the bounces.\\
\subsection{The big billiard}
\begin{figure*}[htbp]
\begin{center}
 \includegraphics[width=0.7\textwidth]{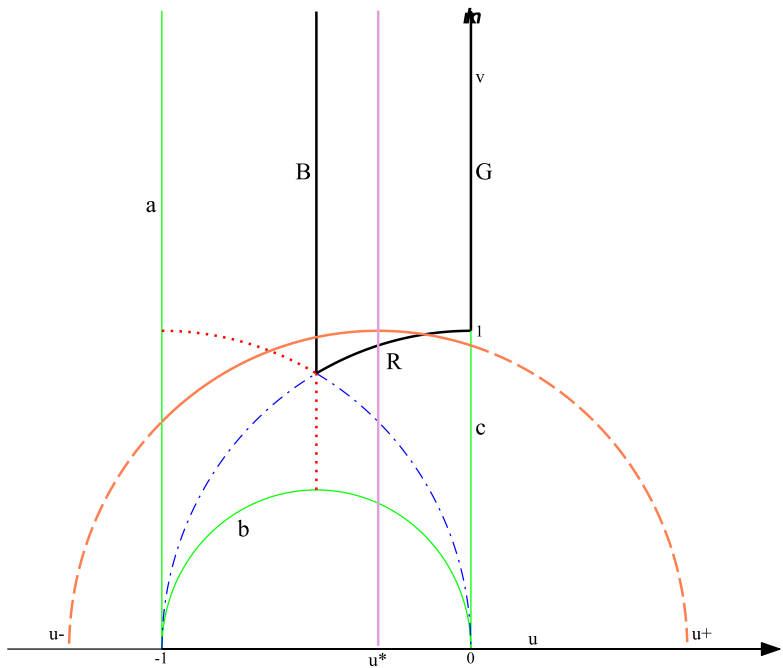}
\caption{\label{figura1} The billiard tables on the UPHP. The big billiard table is delimited by the sides $a$, $b$ and $c$; the subdominant symmetry walls consist of the blue (dashdot) lines bisecting the corners of the big billiard, and of the red (dotted) lines perpendicular to the sides of the billiard, while the dominant symmetry walls define the small billiard table, delimited by the sides $R$, $G$ and $B$. An epoch of the $ba$ type is sketched (orange circle), and is parametrized by the oriented endpoints $u^+$ and $u^-$ of the corresponding geodesics (dashed circle). A generic Poincar\'e surface of section $u^*$, here a generalized geodesics, is represented by the magenta ('vertical') line.}
\end{center}
\end{figure*}

Each approximation to the Kasner solution in the solution of the asymptotic Bianchi IX cosmology is obtained by a different ordering of the triple of Kasner coefficients. The constraints (\ref{eq1}) obeyed by the Kasner coefficients are valid only when the three coefficients are defined within a specific range (\ref{eq2}); when not, a suitable map is considered. The three gravitational walls, in the case of pure gravity, are obtained by fixing an order for the triple of Kasner coefficients. Each bounce on the billiard walls corresponds to a map for the Kasner coefficients in the solution of the Einstein equations, and its representation in terms of the billiard evolution is unique. A trajectory between any two sides of the billiard table is called a Kasner epoch; a collection of epochs taking place in the same corner of the billiard is named a Kasner era. The three corners of the billiard correspond and the two different possible orientations characterizing the first epoch of each era correspond to a symmetry group of order $6$, which, on its turn, reflects the symmetries of the metric tensor. Epochs are named after the sides of the billiard table they join, and eras are named after the first epoch they contain: in Figure \ref{figura1}, an epoch of the $ba$ type is sketched.\\
Several statistical maps and several symmetry-quotienting mechanisms can be defined for the description of the dynamics as a Poincar\'e return map on a suitable surface of section of the billiard table. This way, the continuous dynamics of the billiard is encoded by the statical maps.\\
The succession of Kasner epochs within the same Kasner era is described by the BKL epoch map, and corresponds to the oscillating behavior of two scale factor (related to the corner of the billiard where the oscillations take place), and to the monotonic evolution of the remaining one. The change to the next Kasner era is described by the change of the slope of the derivative of the non-oscillating scale factor, and is described by the CB-LKSKS map. These different maps are defined on different subregions of the restricted phase space.\\
%%%%%%%%%%%%%%%%%%%%%%%%%%%%%%%%%%%%%%%%%%%%%%%%%55
%%%%%%%%%%%%%%%%%%%%%%%%%%%%%%%%%%%%%%%%%%%%%%%%%%%%
%%%%%%%%%%%%%%%%%%%%%%%%%%%%%%%%%%%%%%%%
The big-billiard group (BBG) is obtained from the big billiard table, i.e. a domain defined by the three sides $a$, $b$, $c$,  
\begin{subequations}\label{bbu}
\begin{align}
&a: u=0\\
&b: u=-1\\
&c: u^2+u+v^2=0, 
\end{align}
\end{subequations}
for which bounces against the billiard sides are expressed by the following transformations on the UPHP
\begin{subequations}\label{bbz}
\begin{align}
&Az=-\bar{z},\\
&Bz=-\bar{z}-2,\\
&Cz=-\tfrac{\bar{z}}{2\bar{z}+1}.
\end{align}
\end{subequations}
Eq.'s (\ref{bbz}) are usually referred to as the unquotiented big-billiard map $\mathcal{T}$ for the variable $z=u+iv$ on the UPHP.\\
\\
The quotiented big-billiard map on the UPHP is expressed by the succession of transformations
\begin{equation}\label{BKLz}
z\rightarrow T^{-1}z\rightarrow ... \rightarrow T^{-n+1}z\rightarrow z'\equiv\frac{1}{\bar{z}-n+1}-1\equiv T^{-1}SR_1T^{-n+1}z,
\end{equation}
where the first part of the map is the Kasner quotiented BKL epoch map on the UPHP, while the second part of the map is the Kasner quotiented CB-LKSKS map on the UPHP. In (\ref{BKLz}), the transformations $R_1z=-\bar{z}$, $Tz=z+1$ and $Sz=-1/z$ are defined, where $R_1$ is the only reflection in the maps. The unquotiented big- billiard map and the quotiented big-billiard maps in the restricted phase space are obtained by imposing $v=0$, such that the results of \cite{Damour:2010sz} and \cite{Lecian:2013cxa} are recast.\\
\\
In the restricted phase space, the two-variable maps, consisting of the BKL epoch map, and of the BKL era-transition map, i.e. the CB-LKSKS map, are given by
\begin{equation}\label{BKLupm}
u^\pm\rightarrow u^\pm-1\rightarrow ... \rightarrow u^\pm-(n-1)\rightarrow u^{\pm '}\equiv\frac{1}{u^\pm-n+1}-1,
\end{equation}
where the map acts diagonally on both the variables $u^+$ and $u^-$, but the era-transition map depends on the value of the variable $u^+$ only.\\
\\
The BKL epoch map and the BKL era-transition map, i.e. the CB-LKSKS map, for the statistical variable $u\equiv u^+$ then reads
\begin{equation}\label{BKLup}
u^+\rightarrow u^+-1\rightarrow ... \rightarrow u^+-(n-1)\rightarrow u^{+ '}\equiv\frac{1}{u^+-n+1}-1.
\end{equation}
%%%%%%%%%%%%%%%%%%%%%%%%%%%%%%%%%%%%%%%%%%%%%%%%%%%%%%%%%%%%
\subsection{The small billiard}
The so-called \textit{small billiard} corresponds to the asymptotic limit towards the cosmological singularity in the most general non-homogeneous case: the constraints which have to be satisfied lead one to consider also the symmetry walls, which bisect the angles of the triangular domain of the billiard table, and are perpendicular to the opposite side. As a result, a smaller (with respect to the pure gravitational case) billiard system is obtained, and its boundaries are delimited by (half of) a gravitational wall, and (half of) two symmetry walls.\\
As a result, the side $G$ (Green) is the suitable portion of the $b$ gravitational wall, the side $B$ (Blue) bisects the angle located at $v=\infty$, and the side $R$ (Red) is perpendicular to the side $G$ (and therefore to the side $b$ of the big billiard). The motion of the billiard ball inside the small billiard is still geodesics, and bounces against the sides of the billiard are elastic.\\
\\
An epoch for the small billiard are defined as any trajectory joining any two sides of the small billiard table, and an era for the small billiard is defined as a succession of epochs starting from the side $R$ and ending on the side $R$.\\
The physical meaning of the motion on the small billiard table is traced to the evolution of the scale factors with respect to the suitable time variable. Each bounce on the $G$ side corresponds to a vanishing value for any of the two oscillating scale factor, a bounce on the side $B$ corresponds to an equal value of the two oscillating scale factors, and a bounce on the side $R$ corresponds to an equal value of any of the two oscillating scale factors and the non-oscillating one.\\
As an BKL era is defined by the change of slope of the non-oscillating scale factor, such that, according to the symmetries of the big billiard, it is possible to define an exact connection between the two systems by considering the evolution of the three scale factors and by dividing the reduced phase space according to the different patterns of evolution of the three scale factors. A comparison with the unquotiented dynamics of the big billiard allows one to define a precise map between the two systems. The dynamical equivalence of the two systems is therefore established beyond the geometrical comparison.\\
%%%%%%%%%%%%%%%%%%%%%%%%%%%%%%5
\\
The small billiard is delimited by the sides $G$, $B$, $R$, defined as 
\begin{subequations}\label{smdomain}
\begin{align}
&G: \ \ u=0,\\
&B: \ \ u=-\tfrac{1}{2},\\
&R: \ \ u^2+v^2=1.
\end{align}
\end{subequations}
The transformation that describe the bounces of the billiard ball against the sides of the small billiard table (\ref{smdomain}) are 
\begin{subequations}\label{smalltrasf}
\begin{align}
&R_1(z)=-\bar{z},\\
&R_2(z)=-\bar{z}+1,\\
&R_3(z)=\tfrac{1}{\bar{z}},
\end{align}
\end{subequations}
where no identification among the sides is present, and no symmetry-quotienting mechanisms for the $R$ side of the small billiard table (\ref{smdomain}) is assumed. Eq.'s (\ref{smalltrasf}) are usually referred to as the small-billiard map on the UPHP, According to \cite{Kleinschmidt:2009cv}, reflections on the side $R$ are denominated Weyl reflections, while reflections on the side $B$ and $G$ are called affine reflections, as for the most general classification of all the possible physical transformations qualifying the dynamics of cosmological billiards, up to the number of spacetime dimensions $11=10+1$ for which the phenomenon of chaos qualifies the dynamics of cosmological billiards.\\
Epochs on the small billiard table are defined as any trajectory joining any two walls of the small billiard sides; eras for the small billiard are defined as a succession of epochs starting from the side $R$.\\
%%%%%%%%%%%%%%%%%%%%%%%%%%%%%%%%%%%%%%%%%%%%%%%%%%%
\paragraph{Symmetry-quotienting mechanisms for the small billiard}
The CB-LKSKS map for the small billiard, $t_{CB-LKSKS}$, is defined by two different kinds of transformations, i.e., 
\begin{subequations}\label{tsb}
\begin{align}
&t^{1,2}z=T^{-1}SR_1T^{-n+1}z,\ \ {\rm for} (u^+, u^-)\in S^{1}_{ba} {\rm and} (u^+, u^-)\in S^{2}_{ba},\label{tsb1}\\
&t^{2',3,3'}z=T^{-1}SR_1T^{-n+1}R_3z,\ \ {\rm for} (u^+, u^-)\in S^{2'}_{ba}, (u^+, u^-)\in S^{3}_{ba}, {\rm and} (u^+, u^-)\in S^{3'}_{ba}\label{tsb2}.
\end{align}
\end{subequations}
which act on the subregions of the reduced phase space $S^1_{ba}$, $S^2_{ba}$, $S^3_{ba}$, $S^{2'}_{ba}$ and $S^{3'}_{ba}$ defined as
\begin{subequations}\label{dynreg}
\begin{align}
&S^1_{ba}:\ \ u^-<-\Phi,\ \ u^+>u_\alpha(u^+),\ \ -\Phi<u^-<-1,\ \ u^+>u_\gamma(u^+),\\
&S^2_{ba}:\ \ -\Phi<u^-<-1,\ \ u_\alpha(u^+)<u^+<u_\gamma(u^+),\\
&S^{2'}_{ba}:\ \ u^-<-2,\ \ 0<u^+<u_\alpha(u^+),\ \ -2<u^-<-\Phi,\ \ u_\gamma(u^+)<u^+<u_\alpha(u^+),\\
&S^{3}_{ba}:\ \ -2<u^-<-\Phi,\ \ u_\gamma(u^+)<u^+<u_\beta(u^+),\ \ -\Phi<u^-<-1,\ \ u_\alpha(u^+)<u^+<u_\beta(u^+),\\
&S^{3'}_{ba}:\ \-2<u^-<-1,\ \ 0<u^+<u_\beta(u^+),
\end{align}
\end{subequations}
where the functions
\begin{subequations}\label{functions}
\begin{align}
&u_\alpha(u^+):\ \ u^+=-\tfrac{1}{u^-},\\
&u_\beta(u^+):\ \ u^+=-\tfrac{u^- + 2}{2u^- + 1},\\
&u_\gamma(u^+):\ \ u^+=-\tfrac{u^-+2}{u^-+1},
\end{align}
\end{subequations}
are defined. The 'Golden Ratio' $\Phi$ in (\ref{dynreg}) is approximated as $\Phi\sim 1,618$. The two functions $u_\alpha$ and $u_\beta$ cross at the point $u^-=-\Phi$ and $u^+=\phi$, where the 'Small Golden Ratio' $\phi$ is defined as $\phi+1=\Phi$.\\
The presence of epochs joining different sides of the billiard table in the desymmetrized version of the dynamics is due to the fact that the Poincar\'e return map fr the small billiard is defined in a surface of section different from one defined by an entire gravitational wall of the big billiard table. This way, epoch joining different sides of the billiard are equiparated after mapping them on the regions of the phase space, which correspond to epochs of the same type; this mapping implies the presence of an extra reflection, corresponding to that contained in the pertinent Kasner transformation.\\
Eq.'s (\ref{tsb}) are usually referred to as the quotiented small-billiard map. The difficulty to uniquely define a correspondence between the sequence of epochs and eras between the big billiard and the small billiard leads one to define the unquotiented small-billiard map as the suitable iterations of (\ref{tsb}) which allows a comparison with (\ref{BKLz}). The quotiented small billiard map and the unquotiented small billiard map are obtained by imposing $v=0$ for $z$, such that the results obtained in \cite{Lecian:2013cxa} and in \cite{Damour:2010sz} are recast.\\
\paragraph{The domain of the small billiard and the domain of the big billiard} A desymmetrized small billiard system is obtained as one implied by the predominant walls arising from the constraint of the Einstein field equations in any number ($\le 11$) of dimensions, and its volume is evaluated as FLEIG. The shape of the big billiard domain is one implied by the determination of the congruence subgroup resulting from the union of the number of copies of the small billiard domain necessary to enclose a (hyper) volume having all the vertices places on the absolute of the (generalized) UPHP \cite{Lecian:2013cxa}. 
%%%%%%%%%%%%%%%%%%%%%%%%%%%%%%%%%%%%%%%%%%%%%%%% 
%%%%%%%%%%%%%%%%%%%%%%%%%%%%%%%%%%%%%%%%%%%
\section{Trajectories in Cosmological Billiards\label{section3}}
The symmetries of the billiard table and those of the dynamics allow one to consider suitable symmetry-quotienting mechanisms, such that the six kinds of Kasner eras can be mapped to (a preferred) one. If this preferred one is of the $ba$ type, as depicted in Figure \ref{figura1}, the continued-fraction decomposition of the variable $u^+$ parameterizing the first Kasner epoch of each Kasner era encodes the number of epochs contained in each Kasner era for the evolution of the dynamics, in the symmetry-quotiented version of the dynamics, as  
\begin{equation}\label{periodicu}
u^+_1=n_1-1+x_1\equiv [n_1-1; n_2, n_3, ...],\ \ x_1\equiv [n_2, n_3, ...], 
\end{equation}
where the square brackets denote the fractional part of $u^+$, and $n_1-1$ is its integer part.\\
\\
According to the continued-fraction decomposition of the variable $u^+$, the initial value of the variable $u^+$ can be classified: in the case the continued-fraction decomposition is finite, the value of $u^+$ is rational, and the trajectory will fall into one of the corners of the billiard, which correspond to the cosmological singularity, such that the trajectory is called singular; singular trajectories constitute a countable set. In the case the continued-fraction decomposition is infinite, the value of $u^+$ is irrational, the trajectory will never reach any corner of the billiard, and is called non-singular. Non singular trajectories can be further divided into non-periodic trajectories and periodic trajectories. Furthermore, periodic trajectories are classified into purely periodic trajectories and non-purely-periodic trajectories. Periodic trajectories constitute a countable set (the set of periodic irrationals).\\
The analysis of the continued-fraction decomposition for the variable $u^+$ allows one classify the initial conditions which have to be considered for the solution to the Einstein field equations, and which correspond to the initial conditions for which the cosmological singularity is defined.\\
\\ 
The decomposition (\ref{periodicu}) can be specified for the periodic configuration corresponding a repetition of the sequence $\{\ k\}\ $, where the periodic sequence $k$ is defined as
\begin{equation}\label{eq17}
k\equiv (n_1, n_2, ..., n_k),
\end{equation}
by the notation $u^+_k$
\begin{equation}\label{periodic}
u^+_k=n_1-1+[n_2, n_3, ..., n_k, n_1, ...].
\end{equation}
%%%%%%%%%%%%%%%%%%%%%%%%%%%%%%%%%%%%%%%%%%%%%%%
The invariant form $\omega(u^+, u^-)$,
\be\label{omega}
\omega(u^+, u^-)=\tfrac{1}{2}\frac{du^+\wedge du^-}{(u^+-u^-)^2}
\ee
defines the (non-trivial) area measure on the reduced phase space, and is obtained from the complete symplectic form of Hamiltonian systems evaluated at a fixed energy shell for a Poincar\'e surface of section, and can be expressed as a function of the variables $u^+$ and $u^-$ of the restricted phase space. Probabilities for a succession of eras containing a given number of epochs is obtained as the area of the pertinent subregions of the restricted phase space, according to the non-trivial measure $\omega$.\\
%%%%%%%%%%%%%%%%%%%%%%%%%%%%%%%
\subsection{Sequences of eras and symmetry operations}
For later purposes, it is useful to define, for a sequence $\{\ k \}\ $ Eq. (\ref{eq17}), a symmetry operation on the digits of the sequence $symm( \{\ k \}\ )$, whose digits are denoted by $n_{a_1}$, with $i=1, 2, ..., k$, i.e.
\be\label{symmetry}
symm( \{\ k \}\ )=(n_{a_1}, n_{a_2}, ..., n_{a_k}), i=1, 2, ..., k,
\ee
such that the symmetry operation on a sequence corresponds to a symmetry operation among the generators that compose the sequence, as far as the iterations of the billiard maps are concerned.\\
\\
This way, cyclic permutations on the sequence $\{\ k \}\ $ Eq. (\ref{eq17})are defined by the sequence $cycl( \{\ k \}\ )$, whose digits are denoted by $n_{c_i}$, with $i=1, 2, ..., k$, i.e.
\be\label{cyclic}
cycl( \{\ k \}\ )=(n_{c_1}, n_{c_2}, ..., n_{c_k}), i=1, 2, ..., k,
\ee
such that a cyclic permutation on the generators of the corresponding billiard maps are defined.\\
Similarly, all kinds of permutations (i.e. cyclic permutations between the elements and exchange permutation between any two elements) on the sequence $\{\ k \}\ $ Eq. (\ref{eq17})are defined by the sequence $per( \{\ k \}\ )$, whose digits are denoted by $n_{p_i}$, with $i=1, 2, ..., k$, i.e.
\be\label{permutation}
per( \{\ k \}\ )=(n_{p_1}, n_{p_2}, ..., n_{p_k}), i=1, 2, ..., k,
\ee
and correspond to permutations among the generators of the billiard maps.\\
Particular attention should be paid to the fact that cyclic permutations on the generators of the billiard maps correspond to iterations of the billiard map, as far as periodic orbits are concerned, while exchange permutations among the generators of the map are obtained via the commutators of the generators of the billiard groups, such that the insertion of these commutators within the sequence of transformations that compose the billiard maps can imply a sequence of reflections which does not correspond to the evolution of the scale factors for the solution to the Einstein field equations.\\
\\
It is possible to define also the same symmetry operations on the elements of the sequence $p \{\ k \}\ $ consisting of $p$ times a repetition of a sequence $\{\ k \}\ $, i.e.
\be
p \{\ k \}\ \equiv (n_1, n_2, ..., n_k, n_1, n_2, ...)\ \ p \ \ {\rm times},
\ee
as the symmetry operations acting on the sequence $p \{\ k \}\ $, i.e.
\be\label{symmetry1}
symm(p \{\ k \}\ )=(n_{a_1}, n_{a_2}, ..., n_{a_{pk}}), i=1, 2, ..., pk,
\ee
where the symmetry operation can be specified as a cyclic permutation (\ref{cyclic}) or any permutation (\ref{permutation}), including exchange of two elements.
%%%%%%%%%%%%%%%%%%%%%%%%%%%%%%%%%%%%%%%%%%%%%%
\subsection{Periodic orbits}
Different versions of the dynamics can be schematized according to this formula, such that the limit to a stochastic process, where the dynamics can be considered as almost independent on the initial conditions, acquires different expressions for the one-variable maps and for the two-variable maps in the case of the big billiard and in the case of the small billiard, such that the role of the two variables $u^+$ and $u^-$ can be further understood as fixing the features acquired by cosmological billiards as a large number of iterations of the billiard maps will render the dynamics stochastic.\\
%%%%%%%%%%%%%%%%%%%%%%%%%%%%%%%%%%%%%
\paragraph{Initial conditions, periodic configurations, cyclic permutations, exchange permutations and conjugacy classes}
The relevance of the variables $u^+$ and $u^-$ relies on their properties to describe the initial conditions for the solution to the Einstein field equations in the asymptotic limit to the cosmological singularity.\\
Periodic initial conditions are such that the variable $u^+$ contains and infinite repetition of a periodic sequence of eras. The periodicity of orbits is due to the variable $u^+$. For the two-variable map, periodic conditions are obtained also for the variable $u^-$. Nevertheless, the variable $u^-$ is stable under small perturbations, such that on the one hand, any modification to the initial value of $u^-$ does not perturb the periodicity of the variable $u^+$, and, on the other hand, for non-periodic initial conditions for the variable $u^-$, an infinite number of iterations of the billiard maps is able to bring the non periodic initial value to the periodic one.\\
Periodic orbits are defined by initial values of the statistical variables invariant under the iterations of the billiard maps, i.e. 
\be\label{tu}
\tt{T}u=u,
\ee
where the map $\tt{T}$ corresponds to the composition of matrices whose trace defines the hyperbolic length of the periodic orbit, and $u$ is the statistical variable, for which the periodicity condition is fulfilled, regardless to the symmetry-quotienting mechanism implied. According to the properties of the big-billiard group, i.e the congruence subgroup $\Gamma_2$ of the modular group, this composition is unique. From a physical point of view, there exists only one set of transformation defining the billiard map in Eq. (\ref{tu}) according to the symmetries of the solution to the Einstein filed equations, which are reproduced in the billiard dynamics by fixing the 'labels' of the walls and then considering a symmetry group of order $6$, corresponding to all the possible permutations of the three scale factors $a$, $b$ and $c$ which define the three Kasner parameters in Eq. (\ref{eq1}) and Eq. (\ref{eq2}). From a mathematical point of view, the total number of permutations, i.e. $6$, of the three scale factors corresponds to the number of all the representatives of the three conjugacy subclasses of the permutation group of three elements.\\
Form a mathematical point of view, 'mathematical billiards' are defined by estimating the number of conjugacy classes of a symmetry group.\\
On the contrary, in the following, the strategy followed consists in taking into account all the representatives of the conjugacy classes of a symmetry group, instead of only the number of distinct conjugacy subclass for a given group. This procedure is therefore fully consistent with the symmetries of the Einstein filed equations. Furthermore, the kind of symmetry- operation characterizing the conjugacy subclasses are given a precise physical interpretation within the BKL description, and also a different weight in the sum according to the different statistical maps.\\
\subsection{Periodicity phenomena for the unquotiented big billiard\label{machinery}}
Periodic orbits of the big billiard are a phenomenon which is more complicated than its symmetry-quotiented versions. Given a $m$-periodic orbit of the one-dimensional BKL epoch map
\begin{equation}\label{eq32}
T_{\rm BKL}^m (u^+) = u^+ 
\end{equation}
with 
\begin{equation}\label{eq33}
\sum_{i=1}^{1=k}n_i=m,
\end{equation}
periodic orbits of the big billiard group are given by $mp$ iterations of the unquotiented billiard map $\mathcal{T}$
\begin{equation}\label{mp}
\mathcal{T}^{mp} u=u  ,
\end{equation}
where $m$ is the total number of BKL epochs for which the quotiented big billiard BKL map is periodic, and $p$ is the order of the Kasner transformation for which the new sequence of eras takes place in the correct corner and in the correct orientation.\\
\section{Densities of measure for the billiard maps\label{section4}}
\begin{figure*}[htbp]
\begin{center}
\includegraphics[width=0.7\textwidth]{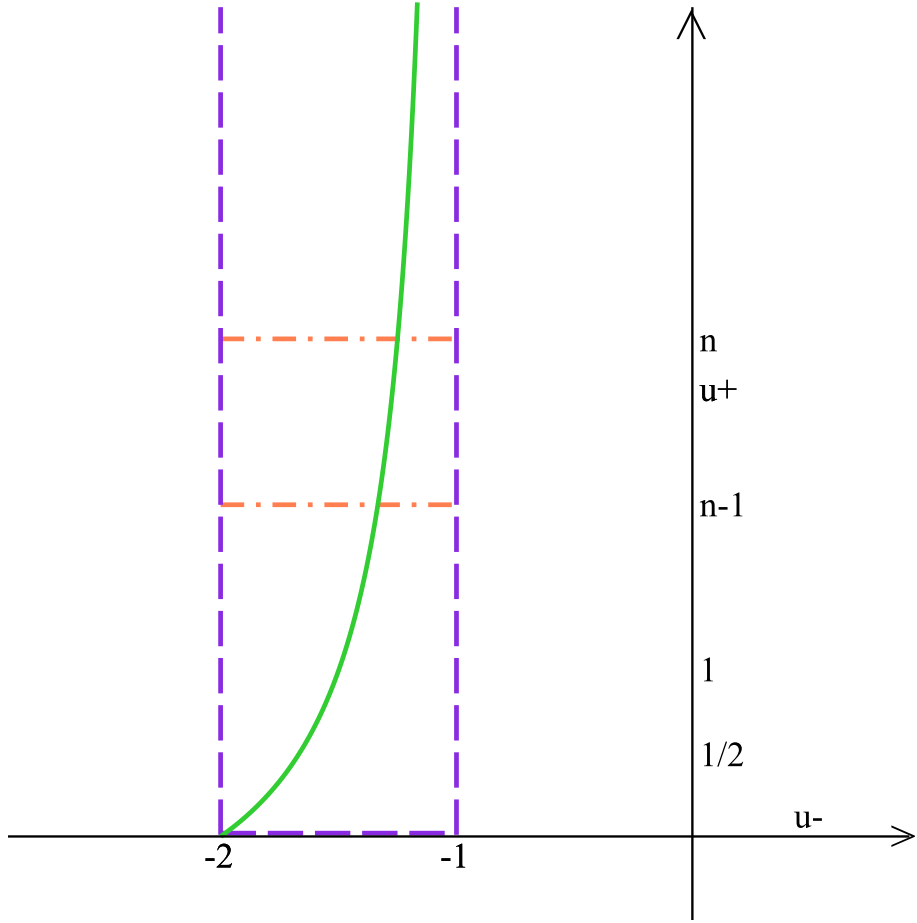}
\caption{\label{figura2} The restricted phase space is parameterized by the variables $u^+$ and $u^-$. Here, the regions available for the quotiented dynamics are illustrated. The subregion corresponding to the Kasner quotiented CB-LKSKS era map for the big billiard is delimited by the violet (dashed) lines. This box is divided by the function $u_\gamma$ (plotted by the green solid line), defined in (\ref{functions}) into two different subregions, defined in (\ref{subregions}), which correspond to the Kasner projection of the dynamical subregions of the restricted phase in (\ref{dynreg}), where the small billiard map is defined, on the UPHP, by a different number of Weyl reflections for the length of the corresponding era in the big billiard. The region $I$ of (\ref{subregions}) is the subdomain enclosed between $u_\gamma$ and $u^+=0$, while the region $II$ is enclosed between $u^+=\infty$ and $u_\gamma$. The orange dashdot line encloses the generic sub-box, for which the probability to find an era containing $n$ epochs is defined, both for the big billiard, and for the small billiard, for the BKL probabilities for the big billiard $P^{BKL}(n)$ (\ref{probb}), and for the BKL probabilities for the small billiard ${\rm p}^{BKL}(n)$ in (\ref{smallprob}), respectively.}
\end{center}
\end{figure*}

The marginalization of the $u^-$ variable from the invariant form $\omega(u^+, u^-)$ leads to the definition of the quantity $W(u^+)$ (after the suitable normalization), which is of the density of the invariant measure for the billiard maps. In the statistical description of discrete variables, the density of invariant measure can be assumed as a probability mass function. For different phenomena, denoted by $\mu$, one starts by defining the normalized density of the invariant measure for the billiard map, i.e.
\begin{equation}\label{densitymu}
W^\mu(u^+)\equiv \tfrac{1}{2A^\mu}\int_{D^\mu(u^-)}du^-\omega(u^+,u^-)
\end{equation}
where the integration regions $D^\mu(u^-)$ and $D^\mu(u^+)$ (below) are the regions of the restricted phase space, which are considered for a specific symmetry-quotiented map, and $A^\mu$ is its area, according to the reduced invariant form $\omega$, i.e.
\be\label{areamu}
A^\mu=\tfrac{1}{2}\int_{D^\mu(u^+)}du^+\int_{D^\mu(u^-)}\omega(u^+,u^-).
\ee
Furthermore, the density of measure $W(u^+)$ defines the corresponding forms $W^\mu(u^+)du^+$ as the marginalization of the variable $u^-$ over different regions of the restricted phase space, which correspond to different physical phenomena.\\
\\
In the case of discrete variables, the normalized invariant density of measure $W(u^+)$ can be taken as a probability mass function for the definition of a probability density for the discrete variables. This is the case of the initial configurations for cosmological billiards which correspond to periodic orbits.\\
A similar characterization can also be adopted for the other countable set of initial conditions, i.e. the countable set of singular trajectories \cite{bel2009}.\\ 
\\
It is worth remarking that the density of measure $W(u^+)$ is linked to the invariant form $\omega(u^+, u^-)$ also via the definition of a normalized invariant measure $W(u^+)du^+$ for the era maps,
\begin{equation}
W(u^+)du^+\equiv\tfrac{du^+}{A^\mu}\int\omega du^-;
\end{equation}
the appearance of measures, defined in this same way for the epoch maps, has been outlined in \cite{Damour:2010sz}.
\paragraph{The big billiard}In the case of the big billiard, in the Kasner-quotiented version of the dynamics, the density of measure invariant under the billiard maps
\begin{equation}\label{eq21}
W(u^+)=\tfrac{1}{\ln 2}\tfrac{1}{(u^++1)(u^++2)},
\end{equation}
where the integration domain for the variable $u^-$ is chosen according to restricted phase space of the big billiard table, i.e. $-2\le u^-\le -1$, and the area of the pertinent region of the restricted phase space $A$ is obtained by integrating the invariant form $\omega$ on the regions of the restricted phase space available for the dynamics for the particular symmetry-quotienting mechanism, where the continued fraction decomposition describes the exact number of epochs in each Kasner era. This region is sketched in Figure \ref{figura2}, where it is delimited by the violet (dashed) lines.\\

\paragraph{The small billiard} For the small billiard, different subcases have to be considered according to the projection of the dynamical subregions of the small billiard restricted phase space to the subdomain of the restricted phase space available for the CB-LKSKS map of the big billiard, in the Kasner symmetry-quotienting of the dynamics. Indeed, the dynamical subregions of the restricted phase space REF, where the small-billiard map acquires different forms, divide the restricted phase space available for the definition of the first epoch of each era according to the different number of reflections contained in the expression of the small-billiard quotiented map on the UPHP.\\
They are evaluated through the general definition (\ref{densitymu}) by specifying it to the dynamical subregions of the restricted phase space (\ref{dynreg}), normalized with (\ref{areamu}). The subregions of the restricted phase space $\mu$ are obtained by the subregions (\ref{dynreg}) by considering the subdomains where the small billiard map for the UPHP implies d different number of reflections, and then by matching the boundaries of the subregions, where the density of invariant measure acquires the same form under the different integration domains $D\mu(u^-)$. In particular, the density of measure for the billiard map can be evaluated according to this different number of reflections, such that one is able to define the densities of measure $W^{I}(u^+)$ accounting for the small-billiard quotiented map (\ref{tsb2}), and a densities of measure $W^{II}(u^+)$ accounting for the small-billiard quotiented map (\ref{tsb1}).\\
As plotted in Figure \ref{figura2}, the restricted phase space available for the quotiented small-billiard map consists of two regions, $\mu=I$ and $\mu=II$, defined as
\begin{subequations}\label{subregions}
\begin{align}
&I:\ \ 0\le u^+\le u_\gamma, -2\le u^-\le -1,\\
&II:\ \ u_\gamma\le u^+\le\infty, -2\le u^-\le-1.
\end{align}
\end{subequations}
whose areas $A^I$ and $A^{II}$ are evaluated according to (\ref{areamu}), 
\be\label{aresa}
A^I\equiv A^{II}\equiv \tfrac{1}{4\ln2}\equiv\tfrac{1}{2}A^{0}:
\ee
the function $u_\gamma$ divides the starting box of the big billiard in the restricted phase space into two subregions of the same area, according to the non-trivial measure $\omega$, which is half the area of the starting box for the big billiard, (here the script $0$ in (\ref{aresa}) refers to the big billiard, and will be omitted), $A^0\equiv A \equiv (\ln2)/2 $.
 On these regions, the densities of measure are evaluated according to the definition (\ref{densitymu}), i.e. 
\begin{subequations}\label{densities}
\begin{align}
&W^I(u^+)=\tfrac{2}{\ln 2}\int^{-1}_{u_\gamma}du^-\omega(u^+,u^-)=\tfrac{1}{2\ln 2}\tfrac{1}{(u^++1)(u^{+2}+2u^++2)},\\
&W^{II}(u^+)=\tfrac{2}{\ln 2}\int_{-2}^{u_\gamma}du^-\omega(u^+,u^-)=\tfrac{1}{2\ln 2}\tfrac{1}{(u^++2)(u^{+2}+2u^++2)}.
\end{align}
\end{subequations}
These results are listed in Table (\ref{table1})
Their dependence on $u^+$ is connected, through the continued-fraction decomposition for the variable $u^+$, to the specific number of epochs in each era of the billiard dynamics, and the presence of these further divisions of the restricted phase space is due to the fact that, for $n=1$, different patterns in the order of crossing of the different scale factors, during their time evolution, is possible. This way, one appreciates that the dynamics of the small billiard, as far as the classification of the initial conditions for periodic orbits in the big billiard is concerned, is much more complicated than for the pure-gravitational case.\\
%%%%%%%%%%%%%%%%%%%%
\begin{table}
\begin{center}
\begin{tabular}{ | l || l | l | l | l | }

\hline
    
$\mu$ & $D^\mu(u^+)$ & $D^\mu(u^-)$ & $A^{\mu}$ & $W^{\mu}(u^+)$\\

  \hline
  \hline
  
  $0$ & $0\le u^+\leq\infty$ & $-2\leq u^-\leq -1 $ & $\tfrac{1}{2}\ln 2$ & $\tfrac{1}{\ln2}\tfrac{1}{(u^++1)(u^++2)}$\\
  
  \hline
  \hline 
  
   $I$ & $0\le u^+\le u_\gamma$ & $-2\le u^-\leq -1$ & $\tfrac{1}{4}\ln 2$ & $\tfrac{1}{2\ln 2}\tfrac{1}{(u^++1)(u^{+2}+2u^++2)}$\\
    
  \hline
  
  $II$ & $u_\gamma\le u^+\le\infty$ & $-2\le u^\leq -1$ & $\tfrac{1}{4}\ln 2$ & $\tfrac{1}{2\ln 2}\tfrac{1}{(u^++2)(u^{+2}+2u^++2)}$ \\

  \hline

\end{tabular}
    %\label{tab:secondlevel}
\end{center}
\caption{\label{table1} The densities of invariant measure $W^\mu(u^+)$, defined in Eq. (\ref{densitymu}) invariant under the billiard maps for the big billiard and for the small billiard are evaluated according to the different subregions of the restricted phase space, depicted in Figure \ref{figura2}, enclosed by the domains specified by the ranges $D(u^+)$ and $D(u^-)$, and normalized according to the corresponding area $A^{\mu}$, defined in Eq. (\ref{areamu}). The case $\mu=0$ corresponds to the big billiard, and the script $0$ is omitted.}
\end{table}
%%%%%%%%%%%%%%%%%%%%%%%%%%%%%%%%%%%%%%%%%%%%%%%%55
%%%%%%%%%%%%%%%%%%%%%%%%%%%%%%%%%%%%%%%%%%%%%%%%%%%%%%%%%
\section{Probabilities for the big billiard\label{section5}}
It is now possible to discuss the phenomenon of stochastization of the dynamics of cosmological billiards as far as the definition of probabilities is concerned, in view of defining the mechanism according to which it is possible to pass from a sum over periodic initial conditions to a sum over the sequences that compose the periodic initial conditions.\\
One starts with recalling that the probability $P(n)$ for an era containing $n$ epochs, within the standard BKL statistical description, in expressed by the area of the pertinent subregion of the restricted phase space, as this area is evaluated according to the invariant measure $omega$ and is invariant under the action of the billiard maps and under the symmetries of the UPHP, of which the Kasner transformations are composed.\\
The probability $P(n)$ reads
\be\label{probbb}
P^{BKL}(n)=\tfrac{1}{\ln2}\int_{-2}^{-1}du^-\int_{n-1}^n\tfrac{du^+}{(u^+-u^-)^2}=\tfrac{1}{\ln 2}\ln\tfrac{(n+1)^2}{n(n+2)}.
\ee 
Accordingly, the probability for a sequence of eras $k$ containing $n_1, n_2, ..., n_k$ epochs each reads
\be\label{probb}
P^{BKL}(n_1, n_2, ..., n_k)=\tfrac{1}{\ln2}\int_{-2}^{-1}du^-\int_{n_1-1+\tfrac{1}{n_2+\tfrac{...}{n_k-1}}}^{n_1-1+\tfrac{1}{n_2+\tfrac{...}{n_k}}}\tfrac{du^+}{(u^+-u^-)^2};
\ee
in general, the probabilities for a different ordering of a given sequence are not equal, as it is straightforward to verify, in the easiest case, for $P(n_1, n_2, n_3)\neq P(n_2, n_1, n_3)$.\\
\\
\subsection{Normalized probabilities for the billiard maps}
It is natural to define, the exact statistical probabilities for cosmological billiards, normalized according to a symmetry operation among the components of a  sequence of eras, from the exact BKL probability (\ref{probb}), as
\be\label{probsymm}
P^{BKL}_{symm}(k)\equiv\tfrac{P(n_1, n_2, ..., n_k)}{\langle P(n_1, n_2, ..., n_k)\rangle_{symm}},
\ee
where the denominator is defined as a sum of the values acquired by the BKL probabilities for a sequence $k$ for all the symmetry operations in (\ref{symmetry}), i.e.
\be\label{denominatorsymm}
\langle P(n_1, n_2, ..., n_k)\rangle_{symm}\equiv\sum_{symm{\{\ k\}\ }}P^{BKL}(k).
\ee
One needs to analyze the role of the variable $u^-$ in the mathematical definition of the Selberg trace formula for cosmological billiards: indeed, a two-variable map has not been taken into account in the definition of the Selberg trace formula for billiards, and not been compared with the one-variable version of the map obtained by the marginalization of the remaining variable.
\subsection{The two-variable map for the big billiard}
It is crucial to remark that, in the case of a two-variable map, even if one assumes a stochastization of the dynamics, considering the probability for a sequence of eras $k=(n_1, n_2, ..., n_k)$ to take place as independent of the order at which the digits of a periodic sequence are considered has to be compared with the understanding that the definition of an initial value for the variable $u^-$ allows one to consider as physical only the cyclic permutations of the digits of the periodic sequence, and not all the possible (exchange) permutations (this is straightforward verified by applying the two-variable billiard map on the two variables: non-physical sequences, obtained by a reordering of the digits of the periodic sequence which are not obtained by a cyclic permutation does not correspond to any solution of the definition of periodicity given in Eq. (\ref{tu}) for the two variables). Similarly, one notes that the variable $u^-$ encodes the past evolution of the billiard dynamics: reordering the digits of the continued fraction decomposition of $u^-$ according to the commutators of matrices according to the group properties they exhibit does not produce any physical evolution for cosmological billiards. Indeed, considering only cyclic permutations as the allowed symmetry operation for the two-variable billiard map corresponds to compare different periodic orbits, which, under the iterations of the billiard maps, are considered equivalent because of their periodic features.\\
The symmetry operation under which Eq. (\ref{probsymm}) has to be specified for the two-variable map are, therefore, cyclic permutations of the periodic sequence $k$. For this, a normalized BKL probability for the two-variable map, with respect to a sequence $k$ of eras, $P^{BKL}_{2-var}(k)$ is defined as
\be\label{probcyc}
P^{BKL}_{2-var}(k)\equiv P^{BKL}_{cycl}(k)\equiv\tfrac{P^{BKL}(n_1, n_2, ..., n_k)}{\sum_{cycl(k)}P^{BKL}(n_1, n_2, ..., n_k)},
\ee
where the BKL probability $P^{BKL}(n_1, n_2, ..., n_k)$ for a sequence of epochs $k$, (\ref{probb}), has been taken into account, the subscript $cycl(k)$ denotes the cyclic permutations of the sequence $k$, as in (\ref{cyclic}), and the denominator of (\ref{probcyc}) is defined through (\ref{denominatorsymm}) as
\be\label{denominatorcycl}
\langle P(n_1, n_2, ..., n_k)\rangle_{cyc}\equiv\sum_{cycl(k)}P^{BKL}(k).
\ee
This definition is consistent with the comparison of the dynamics of different periodic sequences.
\subsection{The one-variable map for the big billiard}
On the contrary, in the assumption of a one-variable billiard map, both for the big billiard and for the small billiard, one sees that \textit{any} permutation of the digits composing the periodic sequence corresponds to a physical sequence of bounces, specified only by the continued fraction decomposition of the variable $u^+$: this way, for the one-variable map, the assumption of a stochastic limit of the properties of the dynamics under a large number of iterations of the billiard maps really corresponds to considering all the elements of all the conjugacy subclasses of a given composition of the billiard maps in Eq. (\ref{tu}).\\
The symmetry operation under which Eq. (\ref{probsymm}) has to be specified for the one-variable map are, therefore, all the permutations of the elements of the periodic sequence $k$. For this, a normalized BKL probability for the one-variable map of a sequence of $k$ eras $P^{BKL}_{1-var}(k)$ is defined as
\be\label{probper}
P^{BKL}_{1-var}(k)\equiv P^{BKL}_{per}(k)\equiv\tfrac{P^{BKL}(n_1, n_2, ..., n_k)}{\sum_{per(k)}P^{BKL}(n_1, n_2, ..., n_k)},
\ee
where the BKL probability $P^{BKL}(n_1, n_2, ..., n_k)$ for a sequence of epochs $k$, (\ref{probb}), has been taken into account, the subscript $per(k)$ denotes all the permutations of the elements of the sequence $k$, and the denominator is defined through (\ref{symmetry}) as
\be\label{denominatorper}
\langle P(n_1, n_2, ..., n_k)\rangle_{per}\equiv\sum_{per{\{\ k\}\ }}P^{BKL}(k).
\ee
\\
This definition of probabilities is consistent with the comparison of different singular trajectories, characterized by different sequences of eras, as, for singular trajectories, the variable $u^-$ is marginalized, and the properties of singular trajectories are encoded in the variable $u^+$ only.
\subsection{Probabilities for the unquotiented dynamics}
A suitable definition for probabilities as far as the unquotiented dynamics of the big billiard is concerned is needed.\\
More in detail, it is necessary to specify these probabilities in a way such that the order $p$ of the Kasner transformation is suitably taken into account, i.e. in the case when periodic phenomena are taken into account for the unquotiented dynamics of the big billiard, i.e.
\be\label{probbpk}
P^{BKL}(p(n_1, n_2, ..., n_k))=P^{BKL}(p(n_1, n_2, ..., n_{pk}))=\tfrac{1}{\ln2}\int_{-2}^{-1}du^-\int_{n_1-1+\tfrac{1}{n_2+\tfrac{...}{n_{pk-1}}}}^{n_1-1+\tfrac{1}{n_2+\tfrac{...}{n_{pk}}}}\tfrac{du^+}{(u^+-u^-)^2}:
\ee
as a result, the order $p$ of the kasner transformation determines a repetition of $p$ times a sequence $K$, and the integration domains for the variable $u^+$ contain therefore the corresponding repetition in the continued-fraction expansion.
\subsection{Probabilities normalized for symmetry operations in the unquotiented dynamics of the big billiard}
For this, by matching the definition of (\ref{probbpk}) with that of the symmetry operations on the elements of a sequence of eras, which allow one to compare different trajectories in the different versions of the billiard maps, the following extension is found
\be\label{probsymmp}
P^{BKL}_{symm}(pk)\equiv\tfrac{P(p(n_1, n_2, ..., n_k))}{\langle P(p(n_1, n_2, ..., n_k))\rangle_{symm}},
\ee
where the order $p$ of the Kasner transformation modifies the definition (\ref{probsymm}) by the prescription to consider $p$ repetitions of the periodic sequence in the evaluation of the numerator (\ref{probb}), and to sum over all the such-specified elements in the denominator, as
\be
\langle P(p(n_1, n_2, ..., n_k))\rangle_{symm}=\sum_{symm\{\ k \}\ }\langle P(p(n_1, n_2, ..., n_k))\rangle_{symm}
\ee
\subsection{Probabilities for the unquotiented two-variable map for the big billiard}
For cyclic permutations of the elements of the periodic sequence, the BKL probability $P^{BKL}(pk)$ is defined as 
\be\label{probcycp}
P^{BKL}_{2-var}(pk)\equiv P^{BKL}_{cycl}(pk)\equiv\tfrac{P^{BKL}(p(n_1, n_2, ..., n_k))}{\langle P(p(n_1, n_2, ..., n_k))\rangle_{cycl}},
\ee
with
\be
\langle P(p(n_1, n_2, ..., n_k))\rangle_{cycl}=\sum_{cycl\{\ k \}\ }\langle P(p(n_1, n_2, ..., n_k))\rangle_{cycl}
\ee
\subsection{BKL probabilities for the one-variable map for the unquotiented big billiard}
The BKL probability for all the permutations of the periodic sequence that constitute the periodic sequence $k$, within the framework of the unquotiented dynamics is generalized as e BKL probability for the one-variable map
\be\label{probperp}
P^{BKL}_{1-var}(pk)\equiv P^{BKL}_{per}(pk)\equiv\tfrac{P^{BKL}(p(n_1, n_2, ..., n_k))}{\langle P(p(n_1, n_2, ..., n_k))\rangle_{per}}.
\ee
with
\be
\langle P(p(n_1, n_2, ..., n_k))\rangle_{per}=\sum_{per\{\ k \}\ }\langle P(p(n_1, n_2, ..., n_k))\rangle_{per}
\ee
%%%%%%%%%%%%%%%%%%%%%%%%%%%%%%%%%%%%%%%%%%%%%%%%%%%%%
\section{A stochastizing BKL dynamics\label{section6}}
The features of the BKL dynamics can be described by BKL probabilities, which define the probability for an exact sequence of eras to take place, or, under the iterations of the billiard maps, as a stochastic process. The definition of BKL probabilities for classes of trajectories, defined by the sets of trajectories which can be compared within the one-variable map or within the two-variable map, allows for the definition of the intermediate regime of the dynamics, where the number of iterations of the billiard map is not still sufficient for the determination of the full stochastic regime, but where the comparison of the trajecotires for the physical interpretation of the one-variable map and of the two-variable map can be done within the framework of a stochastization trend for the dynamics. One therefore has to evaluate the stochastic limit, denoted by $S$, of $P^{BKL}_{symm}(k)$ for the quotiented dynamics of the big billiard, or the stochastic limit of $P^{BKL}_{symm}(pk)$, i.e.
\be
\lim_S \langle \sum_{symm(k)}P^{BKL(k)}\rangle_{symm\{\ k \}\ }
\ee
according to the effects of the intermediate steps of the stochastization process which has to be analyzed within the physical features of the BKL statistical maps. As a result, it is possible to describe the effects of a progressive stochastization of the dynamics by considering a the stochastization of only one particular features of the definition of the BKL probabilities for the one-variable map and for the two-variable map.\\
\\
This intermediate regime of the dynamics is defined by comparing the different trajectories within the same class of trajectories by normalizing the corresponding exact BKL probability by the stochastic limit, of the denominator,
where the limit of the normalization is
\be\label{adesso}
\lim_S \langle \sum_{symm(k)}P^{BKL(k)}\rangle_{symm\{\ k \}\ }\simeq n_{symm}(k)\prod_{i=1}^{i=k}P(n_i),
\ee
where the sum over the BKL probabilities is assumed to almost factorized according to $n_{symm}(k)$ time the exact BKL probability for a single $n_j$, $j=1,..., k$, with $n_symm(k)$ the number of possible symmetry operations on the elements of the sequence $k$.\\
\\
The reason for the description of this intermediate steps in the stochastization of the dynamics is found in the statistical properties of the BKL maps, for which the stochastic limit is reached after a large number of iterations of the billiard map. During this process, it is possible to apply 'external' mechanisms, which, up to some extent, can modify the billiard maps without in principle completely removing the chaotic features of the billiard system: within this framework, the physical information contained in different classes of trajecotires can be analyzed with respect to the common features attributed according to the different billiard map.\\
\\
The next step towards a full stochastization of the dynamics consists in considering the statistical relevance of an exact trajectory according to its exact BKL probability, compared with the other trajectories of the same class, according to the different statistical maps, where the comparison is accomplished via a denominator, which is considered to almost factorize according to the stochastic limit of the BKL probability, as
\be\label{stoclimprobbkl}
\lim_S P^{BKL}(n_1, n_2, ..., n_k)\simeq C_k \prod_{i=1}^{i=k}\tfrac{1}{n_i^2},
\ee
such that 
\be
\langle \sum_{symm(k)}\lim_S P^{BKL(k)}\rangle_{symm\{\ k \}\ }\simeq C_k n_{symm}(k)\prod_{i=1}^{i=k}\tfrac{1}{n_i^2},
\ee
as it is possible, for a non complete stochastization of the process, to complete the almost-factorization of the asymptotic expression of the probabilities as described by a prefactor $C_k$ which takes into account also the length $k$ of the sequence of eras.
%%%%%%%%%%%%%%%%%%%%%%%%%%
\subsection{Stochastizing BKL dynamics for the quotiented big billiard}
By applying the steps towards a completely stochastized dynamics for the quotiented big billiard, the following specifications for the BKL probabilities for a generic symmetry operations which compose the iterations of the billiard maps are obtained
\be
\lim_S P^{BKL}_{symm}(k)\equiv\tfrac{P^{BKL}(k)}{n_{symm}(k)P^{BKL}(k)}
\ee
where each summand in the normalization is assumed to have the same expression.\\
\\
By assuming that each summand in the normalization consists of the product of the probabilities for each era considered as independent of the initial conditions, the following decomposition holds for the stochastizing process
\be
\lim_S P^{BKL}_{symm}(k)\simeq\tfrac{P^{BKL}(k)}{n_{symm}(k)\prod_{j=1}^{j=k}P^{BKL}(k)},
\ee
while taking into account the asymptotic limit of each independent probability for the lengths $n_j$ of the eras in the product leads to the following most general result
\be
\lim_S P^{BKL}_{symm}(k)\simeq\tfrac{P^{BKL}(k)}{n_{symm}(k)C_(k)\prod_{i=1}^{i=k}\tfrac{1}{n_i^2}},
\ee
where, within the framework of a not completely stochastized dynamics, each expansion in the asymptotic limit can still be considered as dependent of $k$.
%%%%%%%%%%%%%%%%%%%%%%%%%%%%%%%%%%%%%%%%%%%%%%%%%%%%
\subsection{Stochastizing BKL dynamics for the two-variable map of the quotiented big billiard}
The two-variable map allows one to compare two trajecotires considered as different stages of the iterations of the billiard maps on the same trajectory, such that, during the process, the dynamics is hypothesized to partially stochastize. In the case the full stochastized regime has not been taken into account yet, the different steps are described by the successive stochastic approximations as 
\be
\lim_S P^{BKL}_{2-var}(k)\equiv\tfrac{P^{BKL}(k)}{(k)P^{BKL}(k)},
\ee
where $k\equiv n_{cycl}(k)$; the next step is assumed to render the probability for a $n_j$, $j=1, ..., k$ era as independent of the initial conditions, i.e.
\be
\lim_S P^{BKL}_{cycl}(k)\simeq\tfrac{P^{BKL}(k)}{(k)\prod_{j=1}^{j=k}P^{BKL}(k)},
\ee
and considering the asymptotic limit of each product in the summand of the denominator leads
\be
\lim_S P^{BKL}_{cycl}(k)\simeq\tfrac{P^{BKL}(k)}{kC_(k)\prod_{i=1}^{i=k}\tfrac{1}{n_i^2}},
\ee
where the most general non-stochastized dependence is taken onto account.
%%%%%%%%%%%%%%%%%%%%%%%%%%%%%
\subsection{Stochastizing BKL dynamics for the one-variable quotiented big billiard}
The one-variable quotiented big billiard map allows to express the stochastizing process of the BKL dynamics by codifying the properties of a class of trajectories independently of the 'memory' of the system on the 'past' evolution encoded i the variable $u^-$, by approximating the BKL probability as
\be
\lim_S P^{BKL}_{1-var}(k)\equiv\tfrac{P^{BKL}(k)}{(k!)P^{BKL}(k)}
\ee
where each term in the denominator is hypothesized to acquire the same partial stochastic limiting expression.\\
\\
By considering such a partially stochastic limiting expression as independent of the initial conditions, the following intermediate expression describes the next step of the stochastization process as
\be
\lim_S P^{BKL}_{1-var}(k)\simeq\tfrac{P^{BKL}(k)}{(k!)\prod_{j=1}^{j=k}P^{BKL}(k)};
\ee
the most generalized result for this analysis is given by the assumption of the asymptotic expression for each factorized product as depending on the sequence $k$ as
\be
\lim_S P^{BKL}_{1-var}(k)\simeq\tfrac{P^{BKL}(k)}{(k!)C_(k)\prod_{i=1}^{i=k}\tfrac{1}{n_i^2}},
\ee
where the iterations of the billiard maps are supposed to have rendered the BKL dynamics not yet completely stochastic.
\subsection{Stochastizing BKL dynamics for the unquotiented big billiard}
The analysis of the several steps which define the complete stochastization process for the unquotiented big billiard are sketched in the following.\\
\\
The first step in the stochastization process of the dynamics of the unquotiented big billiard is described by considering the behavior of the BKL probability for a repetition of $p$ times a sequence $k$ in the normalization of the denominators for a given statistical map, such that
\be
\lim_S P^{BKL}(p(n_1, n_2, ..., n_k))\rightarrow p P^{BKL}(n_1, n_2, ..., n_k),
\ee
for which the BKL probability for a statistical map is expressed by normalizing the exact BKL probability for a sequence of eras by this limit of the denominators as
\be
\lim_S P^{BKL}_{symm}(p(n_1, n_2, ..., n_k))\simeq\tfrac{P^{BKL}(p(n_1, n_2, ..., n_k))}{\langle p P^{BKL}(n_1, n_2, ..., n_k)\rangle_{symm}},
\ee
such that it is possible to compare the stochastizing dynamics of the unquotiented big billiard with that of the quotiented big billiard.\\
\\
A further specification of the BKL probabilities normalized according to a symmetry operation, which qualifies the physical interpretation of the one-variable map and that of the two-variable map, is expressed by the limit of the BKL probabilities for the sequence $k$ as
\be
\langle p P^{BKL}(n_1, n_2, ..., n_k)\rangle_{symm}\simeq pn_{symm}(k)\prod_{j=1}^{j=k}P^{BKL}(n_k),
\ee 
i.e. by considering the probability for a sequence of eras to take place as the product of the exact BKL probabilities for the corresponding eras $n:j$ to take place independently of the initial conditions expressed by $u^+$, such that the following expression for the limit holds
\be
\lim_S P^{BKL}_{symm}(pk)\simeq\tfrac{P^{BKL}(pk)}{pn_{symm}(k)\prod_{j=1}^{j=k}P^{BKL}(n_k)}.
\ee
An even further step in the analysis of the transformations of the features of the BKL dynamics under the successive iterations of the billiard map is considering the relevance of a sequence of eras within a particular version of the BKL maps as normalized according to fact that BKL probabilities for $p$ times the repetition of a sequence $k$ of eras almost factorize as
\be\label{stoclimprobbklunq}
P^{BKL}(p(n_1, n_2, ..., n_k))\simeq p\prod_{i=1}^{i=k}P(n_i)\simeq pC' \prod_{i=1}^{i=k}\tfrac{1}{n_i^2},
\ee
where $C'$ is the suitable coefficient. These definition is therefore specified for the particular BKL statistics of the unquotiented big billiard, where the order $p$ of the Kasner transformation needed for the comparison with the unquotiented dynamics is taken into account, and where the repetition of $p$ time a periodic sequence corresponding to a closed geodesics for the quotiented dynamics is interpreted as the physical periodic trajectory of the billiard ball on the big billiard table. The corresponding limit of the BKL probability for a given symmetry operation on the generators of the reflections that compose the billiard map acquires therefore the form
\be
\lim_S P^{BKL}_{symm}(pk)\simeq\tfrac{P^{BKL(pk)}}{pn_{symm}(k)C'\prod_{i=1}^{i=k}\tfrac{1}{n_i^2}}
\ee
where the factorization of the stochastic limit of the BKL probabilities is assumed as independent of the length of the sequence $k$ and of the order $p$. More in general, it is possible to keep these precise dependence by writing
\be
\lim_S P^{BKL}_{symm}(pk)\simeq\tfrac{P^{BKL(pk)}}{pn_{symm}(k)C'_{p,k}\prod_{i=1}^{i=k}\tfrac{1}{n_i^2}},
\ee
where this occurrence is traced in the precise expression of the BKL probabilities for the sequence $pk$ of eras.
\subsection{Stochastizing BKL probabilities for the one-variable map of the unquotiented big billiard}
Applying the steps of the stochastization process for the very complicated BKL dynamics to the one-variable BKL statistical maps allows for a comparison for the same steps of the stochastization trend obtained in the quotiented version of the dynamics, with
\be
\lim_S P^{BKL}_{1-var}(pk)\simeq \tfrac{P^{BKL}(pk)}{\langle P^{BKL}(pk)\rangle_{per}}\simeq \tfrac{P^{BKL}(pk)}{p\langle P^{BKL}(k)\rangle_{per}}.
\ee
Considering the probability for the occurrence of each BKL era composed of $n-j$ epochs, $j=1, ..., k$ as independent of its position in the sequence $k$ for the normalization which defines the BKL probabilities for this particular map brings
\be
\lim_S P^{BKL}_{1-var}(pk)\simeq \tfrac{P^{BKL}(pk)}{pk! P^{BKL}(k)},
\ee
and considering the stochastic limit for the BKL probabilities allows one to recover the following expression
\be
\lim_S P^{BKL}_{1-var}(pk)\simeq \tfrac{P^{BKL}(pk)}{p(k!)C'_{p,k}\prod_{j=1}^{j=k}\tfrac{1}{n_j^2}},
\ee
where the most detailed expression of the prefactor $C'$ is considered. In all these expressions, the presence of the prefactor $k!$ is typical of the BKL one-variable map.
\subsection{Stochastizing BKL probabilities for the two-variable map of the unquotiented big billiard}
By specifying these steps to the BKL probabilities for the two-variable map, it is straightforward to verify that its comparison with the unquotiented dynamics reads
\be
\lim_S P^{BKL}_{2-var}(pk)\simeq \tfrac{P^{BKL}(pk)}{\langle P^{BKL}(pk)\rangle_{cycl}}\simeq \tfrac{P^{BKL}(pk)}{p\langle P^{BKL}(k)\rangle_{cycl}}.
\ee
Considering the BKL probabilities for $p$ times the repetition of the sequence $k$, composed of the product of independent probabilities for the eras containing $n-j$ epochs, $j=1, ..., k$ in the normalization yields
\be
\lim_S P^{BKL}_{2-var}(pk)\simeq \tfrac{P^{BKL}(pk)}{pk P^{BKL}(k)},
\ee
and considering the stochastic limit for the BKL probabilities for each era brings the expression
\be
\lim_S P^{BKL}_{2-var}(pk)\simeq \tfrac{P^{BKL}(pk)}{pkC'_{p,k}\prod_{j=1}^{j=k}\tfrac{1}{n_j^2}},
\ee
where the most precise determination of the prefactor $C'$ is considered.
\section{Stochastization of the dynamics of the big billiard\label{section7}}
It is necessary to comment that the definitions of the BKL probabilities for different symmetry operations on the elements of the sequences of eras, according to the physical interpretation of the one-variable map and of the two-variable map, allow one to compare the statistics implied by these expressions and their physical meaning for the evolution of cosmological billiards, in comparison with the definition of conjugacy subclasses used within the framework of 'mathematical billiards'.\\
Indeed, for mathematical billiards, a repetition of the same periodic sequence for the dynamics of a billiard system is ruled out of any particular relevance, as implied by the definition of primitive orbits. Within the framework of cosmological billiards, on the contrary, the repetition of $p$ times a periodic sequence, with $p=1, 2, 3$, is understood within the framework of the definition of periodicity for the unquotiented big billiard dynamics.\\
\\
The iterations of the billiard maps render the dynamics stochastic. In the limit of stochastic features of the dynamics, denoted by the script $S$, after a suitably large number of iterations of the billiard map, Eq. (\ref{probb}) factorizes according to 
\be
\lim_S P^{BKL}(n)\simeq\tfrac{1}{\ln2}\tfrac{1}{n^2}.
\ee
while 
\be
\lim_S P^{BKL}(k)\simeq C\prod_{j=1}^{j=k}\tfrac{1}{n_j^2}
\ee
where $C$ is the suitable coefficient.\\
While Eq. (\ref{probbb}) depends on the order $i=1, ..., k$ of the eras, Eq. (\ref{probb}) does not, nor does the prefactor $C$ any more, which, within the incomplete stochastic description, was allowed to depend on $k$ still as $C_k$. In the stochastic limit, therefore, the probability for a succession of eras to take place is independent of the order of digits composing the sequence $k$, as far as cosmological billiards are concerned.\\
The (normalized according to s particular symmetry operation) probabilities for stochastic processes of cosmological billiards become therefore independent of the dynamics of cosmological billiards, but depend only on the symmetry operation.\\
\subsection{Stochastization of the dynamics in the quotiented big billiard}
For a stochastic process, the limit of the probability $P^{S}(k)$ (where the superscript $S$ refers to the limit to a stochastic process) is described by (\ref{stoclimprobbkl}).\\
For this case, the limit of a probability according to the hypothesis of a stochastic process is obtained as
\be\label{Markovsymm}
P^{S}_{symm}(k)=\tfrac{P^{S}(k)}{\langle P^{S}(k)\rangle_{symm}}
\ee 
where the subscript $symm (k)$ implies a sum over the symmetry transformations of the matrices that compose the billiard map, and which correspond to the digits contained in the periodic sequence $k=(n_1, n_2, ..., n_k)$, and $\langle P^{S}(k)\rangle$ is the average of $P^{S}(k)$, such that.
\be\label{Markovmean}
\langle P^{S}(k)\rangle_{symm}\approx \sum_{symm(k)}P^S(k)\approx n_{symm}(k) C \prod_{j=1}^{j=k}\tfrac{1}{n_i^2},
\ee
where $n_{symm}(k)$ stands for the number of elements obtained from the symmetry operation.\\
\\
As a result, the procedure adopted is be able to uncover the features of the evolution of the scale factors when different initial conditions are considered.\\
Moreover, one is able to appreciate the different physical explanations for the different conjugacy subclasses of a given matrix.
\subsection{Stochastic limit for the two-variable maps of the quotiented big billiard}
For the case of cyclic permutations of the digits composing the periodic sequence, one obtains
\be\label{Markovcyc}
P^{S}_{2-var}(k)\equiv P^{S}_{cycl}(k)=\tfrac{P^{S}(k)}{\langle P^{S}(k)\rangle_{cycl}}
\ee 
where the subscript $cycl (k)$ implies a sum over the cyclic permutations of the digits contained in the periodic sequence $k=(n_1, n_2, ..., n_k)$, and $\langle P^{S}(k)\rangle$ is the normalizing mean of $P^{S}(k)$, such that.
\be\label{Markovcycnorm}
\langle P^{S}(k)\rangle_{cycl}\approx \sum_{cycl(k)}P^S(k)\approx k \prod_{j=1}^{j=k}P(n_j),
\ee
for $n_{cycl}(k)=k$.
\subsection{Stochastic limit for the one-variable maps of the quotiented big billiard}
When all the permutations are considered, the pertinent stochastic normalized probability for the comparison of sequences of eras within the one-variable map is expressed by
\be\label{Markovper}
P^{S}_{1-var}(k)\equiv P^{S}_{per}(k)=\tfrac{P^{S}(k)}{\langle P^{S}(k)\rangle_{per}}
\ee 
and Eq. (\ref{Markovmean}) rewrites
\be\label{markovlimper}
\langle P^{S}(k)\rangle_{per}\approx \sum_{per (k)}P^S(k)\approx k! \prod_{j=1}^{j=k}P(n_j),
\ee
with $n_{per}(k)=k!$.\\ 
\subsection{Stochastization of the dynamics in the unquotiented big billiard}
The BKL probabilities for the different statistical maps are defined according to the symmetry operation among the sequence of eras which have to be compared in the stochastic limit, by defining the stochastic limit of BKL probabilities normalized according to a symmetry operation
\be
\lim_S P^{BKL}_{symm}(pk)=\lim_S \tfrac{P^{BKL}(pk)}{\langle P^{BKL}(pk)\rangle}
\ee
\\
where the denominators admit the limit
\be
\lim_S \langle P(p(n_1, n_2, ..., n_k))\rangle_{symm}\simeq n_{symm}(pk) \lim_S P^{BKL}(n_1, n_2, ..., n_k),
\ee
with the degeneracy prefactor $n_{symm}(pk)$ defined by $k$ and $p$.
\subsection{Stochastization of the dynamics for the two-variable map in the unquotiented big billiard}
The stochastic limit of the BKL probabilities for the two-variable statistical map in the unquotiented version of the dynamics, within the stochastic limit, i.e. within the regime reached after a large number of iterations of the billiard map, is defined as 
\be
\lim_S P^{BKL}_{2-var}(pk)\equiv \lim_S P^{BKL}_{cycl}(pk)=\lim_S \tfrac{P^{BKL}(pk)}{\langle P^{BKL}(pk)\rangle_{cycl}}
\ee
with the corresponding degeneracy prefactor
\be
n_{cycl}(pk)=pk,
\ee
such that
\be
\lim_S P^{BKL}_{2-var}(pk)\simeq \tfrac{1}{pk}.
\ee
\subsection{Stochastization of the dynamics for the one-variable map in the unquotiented big billiard}
The stochastic limit of the BKL probabilities for the one-variable statistical map in the unquotiented version of the dynamics, i.e. within the regime reached after a large number of iterations of the billiard map, is defined as 
\be
\lim_S P^{BKL}_{1-var}(pk)\equiv \lim_S P^{BKL}_{per}(pk)=\lim_S \tfrac{P^{BKL}(pk)}{\langle P^{BKL}(pk)\rangle_{per}}
\ee
with the corresponding degeneracy prefactor
\be
n_{per}(pk)=(pk)!,
\ee
such that
\be
\lim_S P^{BKL}_{1-var}(pk)\simeq\tfrac{1}{(pk-1)!}.
\ee
\section{Probabilities for the Small Billiard\label{section8}}
\begin{figure*}[htbp]
\begin{center}
 \includegraphics[width=0.7\textwidth]{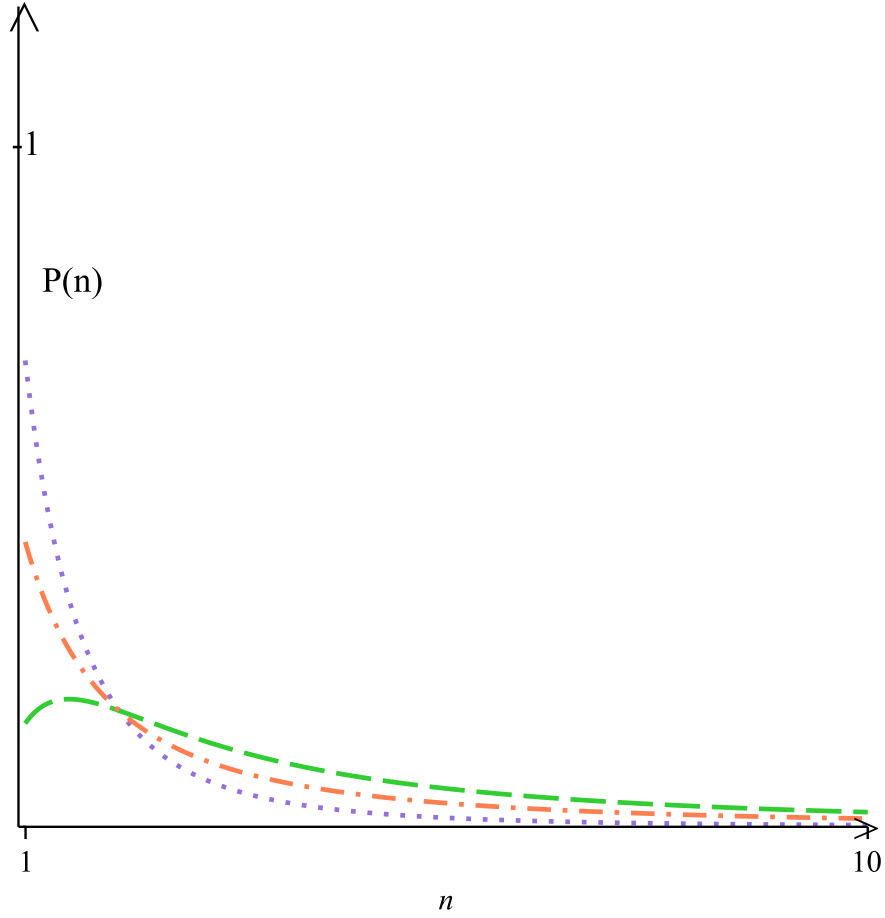}
\caption{\label{figura3} BJK probabilities for the big billiard and for the small billiard: the BKL probabilities $P^{BKL}(n)$ for the big billiard are plotted by the dashdot line;the BKL probabilities ${\rm p}^{BKL}_{I}(n)$ for the map (\ref{tsb2}) of the small billiard are plotted by the dashed line; the BKL probabilities ${\rm p}^{BKL}_{II}(n)$ for the BKL map (\ref{tsb1}) are plotted by the dotted line. The particular features of the BKL probabilities ${\rm p}^{BKL}_{I}(n)$ for small $n$ are illustrated.}
\end{center}
\end{figure*}

it is possible to define probabilities for the small billiard table as well, within the framework of the BKL map for the small billiard on the UPHP, within the perspective of a comparison with the big billiard dynamics.\\
For this, according to the continued-fraction decomposition of the variable $u^+$, and according to the correspondence between the number of epochs in each era of the big billiard and the corresponding map for the small billiards, it is possible to define BKL probabilities for the small billiard.\\
The probability for a $n$-epoch era to take place, within the small billiard map (\ref{tsb}) is defined by the BKL probability for the small billiard ${\rm p}^{BKL}_\mu(n)$, according to the region $\mu$ of the restricted phase space, as the integral of the density of measure $W^{\mu}(u^+)$ aver the pertinent range of $u^+$ as 
\be
{\rm p}^{BKL}_\mu(n)\equiv\int_{n-1}^{n}W^{\mu}(u^+)du^+.
\ee
According to the region $\mu$ of the restricted phase space, one obtains the probability for an $n$ epoch era of the big billiard to take place, and to be characterized by the initial condition $u^+$ and $u^-$, where the variable $u^-$ is defined within a region $\mu$ of the restricted phase space of the small billiard, characterized by a certain number of reflections by the small billiard quotiented map (\ref{tsb}), i.e.
\begin{subequations}\label{smallprob}
\begin{align}
&{\rm p}^{BKL}_{I}(n)\equiv\int_{n-1}^{n}W^{I}(u^+)du^+=\tfrac{1}{2\ln 2}\ln\tfrac{(n+1)^2(n^2+1)}{n^2(n^2+2n+2)}\\
&{\rm p}^{BKL}_{II}(n)\equiv\int_{n-1}^{n}W^{II}(u^+)du^+=\tfrac{1}{2\ln 2}\ln\tfrac{(n+1)^2(n^2+2n+2)}{n^2(n^2+1)(n+2)^2}
\end{align}
\end{subequations}
The results of the definitions (\ref{smallprob}) are reported in Table (\ref{table2}).\\
%%%%%%%%%%%%%%%%%%%%%%%%%%%%
\begin{table}
\begin{center}
    \begin{tabular}{ | l | l | l |}
    \hline
     $P^{BKL}(n)=\tfrac{1}{\ln 2}\ln\tfrac{(n+1)^2}{n(n+2)}$ & ${\rm p}^{BKL}_{I}(n)=\tfrac{1}{2\ln 2}\ln\tfrac{(n+1)^2(n^2+1)}{n^2(n^2+2n+2)}$ & ${\rm p}^{BKL}_{II}(n)=\tfrac{1}{2\ln 2}\ln\tfrac{(n+1)^2(n^2+2n+2)}{n^2(n^2+1)(n+2)^2}$   \\ \hline  
    \end{tabular}
\end{center}
\caption{\label{table2} The expression of BKL probabilities for the occurrence of eras containing $n$ epochs for cosmological billiards. The BKL probabilities $P^{BKL}$ refer to the big billiard, and are recalled in Eq. (\ref{probbb}), where the script $0$ is omitted. The BKL probabilities ${\rm p}^{BKL}_{\mu}(n)$ refer to the small billiard, and are defined in Eq.'s (\ref{smallprob})}
\end{table}
%%%%%%%%%%%%%%%%%%%%%%%%%%%%%%%
One sees that the sum of the two probabilties equals the probability $P^{BKL}(n)$ defined for the big billiard, 
\be
{\rm p}^{BKL}_{I}(n)+{\rm p}^{BKL}_{II}(n)=P^{BKL}(n).
\ee
Furthermore, because of the features of $u_\gamma$, in the limit for a large number of epochs in an era, $n>>1$, the two probabilities admit the expansion
\begin{subequations}\label{smallprobexpansion}
\begin{align}
&{\rm p}^{BKL}_{I}(n)\simeq \tfrac{1}{2\ln2}\left(\tfrac{1}{n^3}+\mathcal{O}\left(\tfrac{1}{n^4}\right)\right),\\
&{\rm p}^{BKL}_{II}(n)\equiv\simeq \tfrac{1}{2\ln2}\left(\tfrac{1}{n^2}-\tfrac{2}{n^3}+\mathcal{O}\left(\tfrac{1}{n^4}\right)\right),
\end{align}
\end{subequations}
such that
\be
\lim_{n>>1}\left({\rm p}^{BKL}_{I}(n)+{\rm p}^{BKL}_{II}(n)\right)\simeq\tfrac{1}{\ln2}\left(\tfrac{1}{n^2}-\tfrac{1}{2n^3}\mathcal{O}\left(\tfrac{1}{n^4}\right)\right),
\ee
such that the asymptotic behavior is recast at all orders. For this, and also by observing that the function $u_\gamma$ tends asymptotically, for large $n$, to $u^-\rightarrow-1$ for $u^+=n>>1$, i.e. that, the larger $n$, the smaller the area of corresponding subdomain of the sub box, it is straightforward compare the two probabilties for large $n$ as
\be
{\rm p}^{BKL}_{II}(n)\simeq P^{BKL}_{I}(n)+\eta(n),
\ee 
with
\be
\eta(n)\equiv\tfrac{1}{2\ln 2}\tfrac{1}{n^3}.
\ee
Indeed, for large $n$, the probability ${\rm p}^{BKL}_{I}(n)$ becomes negligible with respect to the probability ${\rm p}^{BKL}_{II}(n)$, such that the asymptotic limit for large $n$ is close to that of the big billiard.\\
\\
The probabilities ${\rm p}^{BKL}_{I}(n)$ and ${\rm p}^{BKL}_{II}(n)$ for the small billiard are compared with the probabilities $P^{BKL}(n)$ in Figure \ref{figura3}.\\
%%%%%%%%%%%%%%%%%%%%%%%%%%%%%%%%%%%%%%%%%%
%%%%%%%%%%%%%%%%%%%%%%%%%%%%%%%%%%%%%%%%%%%%%%%%%%%%%%%
\\
On the contrary, the main differences of the dynamics of the small billiard with respect to that of the big billiard are exhibited for small values of $n$, and, in particular, for $n\sim1$ the main differences are expected to outline the peculiarities of the small billiard dynamics.\\
\\
Following the comparison of the dynamics of the big billiard with that of the small billiard, it is also possible to define the probability for a succession of eras $\{\ k \}\ $ to take place. Accordingly, the BKL probabilities ${\rm p}^{BKL}_{\mu}(k)$ are defined, for the needed specifications on the elements of the sequence $k$, as 
\begin{subequations}\label{smallprobsequence}
\begin{align}
&{\rm p}^{BKL}_{I}(n_1, n_2, ..., n_k)\equiv\int_{\tfrac{1}{n_2+\tfrac{...}{n_k-1}}}^{\tfrac{1}{n_2+\tfrac{...}{n_k}}}W^{I}( u^+)du^+,\\
&{\rm p}^{BKL}_{II}(n_1, n_2, ..., n_k)\equiv\int_{\tfrac{1}{n_2+\tfrac{...}{n_k-1}}}^{\tfrac{1}{n_2+\tfrac{...}{n_k}}}W^{II}(u^+)du^+,\\
\end{align}
\end{subequations}
where the integration domains are specified according to the definitions of the densities of measure $W^{\mu}(u^+)$ Eq.'s (\ref{densitymu}) specified for the small billiard.
%%%%%%%%%%%%%%%%%%%%%%%%%%%%%%%%%%
\subsection{Probabilities normalized for different statistical maps in the quotiented dynamics of the small billiard}
For a discussion of the dynamics of the small billiard, and for a comparison with the same behavior of the big billiard, the effect of the iterations of the small billiard maps can be discussed by defining BKL probabilities for the small billiard map, normalized according to a suitable symmetry operation of the elements that constitute a sequence $k$, whose repetition defines a sequence of eras in the big billiard table, such that the corresponding trajectory for the small billiard table is defined via the suitable map.\\
According to s suitable symmetry operations, one defines the probability ${\rm p}^{BKL}_{\mu  symm}(k)$ according to the subregions $\mu$ of the restricted phase space available for the description of the dynamics under the small billiard map (\ref{tsb}), according to the number of reflections contained in the small billiard map (\ref{tsb}):
\begin{subequations}\label{smallprobsymm}
\begin{align}
&{\rm p}^{BKL}_{I symm}(k)\equiv \tfrac{{\rm p}^{BKL}_{I}(k)}{\langle {\rm p}^{BKL}_{I}(k)\rangle_{symm}},\\
&{\rm p}^{BKL}_{II symm}(k)\equiv \tfrac{{\rm p}^{BKL}_{II}(k)}{\langle {\rm p}^{BKL}_{II}(k)\rangle_{symm}}.
\end{align}
\end{subequations}
The denominators that normalize the definition (\ref{smallprobsymm}) according to a particular symmetry operation are defined according to the sequences allowed for each region $\mu$, as
\begin{subequations}\label{smallprobdenomsymm}
\begin{align}
&\langle {\rm p}^{BKL}_{I}(k)\rangle_{symm}\equiv\sum_{symm(k)}{\rm p}^{BKL}_{I}(k)\\
&\langle {\rm p}^{BKL}_{I}(k)\rangle_{symm}\equiv\sum_{symm(k)}{\rm p}^{BKL}_{II}(k).
\end{align}
\end{subequations}
\\
\subsection{The two-variable map for the quotiented small billiard}
By applying the statistical meaning of the two-variable map to the small billiard, i.e. by considering that a comparison of cyclic permutations of the elements of a sequence whose repetition defines an orbit of the big billiard, one finds that comparing the BKL probabilties for the small billiard at different cyclic permutations of the number $n_k$ corresponds to comparing the dynamics of periodic orbits at different iterations of the two-variable map.\\
As a result, the following generalizations of the previous definition are found
\begin{subequations}\label{smallprobcycl}
\begin{align}
&{\rm p}^{BKL}_{I 2-var}(k)\equiv {\rm p}^{BKL}_{I  cycl}(k)\equiv \tfrac{{\rm p}^{BKL}_{I}(k)}{\langle {\rm p}^{BKL}_{I}(k)\rangle_{cycl}},\\
&{\rm p}^{BKL}_{II 2-var}(k)\equiv{\rm p}^{BKL}_{II cycl}(k)\equiv \tfrac{{\rm p}^{BKL}_{II}(k)}{\langle {\rm p}^{BKL}_{II}(k)\rangle_{cycl}},
\end{align}
\end{subequations}
where the denominators of (\ref{smallprobcycl}) are defined as
\begin{subequations}
\begin{align}
&\langle {\rm p}^{BKL}_{I}(k)\rangle_{cycl}\equiv\sum_{cycl(k)}{\rm p}^{BKL}_{I}(k)\\
&\langle {\rm p}^{BKL}_{II}(k)\rangle_{cycl}\equiv\sum_{cycl(k)}{\rm p}^{BKL}_{II}(k).
\end{align}
\end{subequations}
%%%%%%%%%%%%%%%%%%%%%%%%%%%%%%%%%%%%%%%%%%%%%%%%%%%%%%%%
\subsection{The one-variable map for the quotiented small billiard}
Similarly to the previous results, it is straightforward to consider that, when only the one-variable map is taken into account, reshuffling the element of a sequence according to any kind of permutations corresponds to compare different trajectories, originated by different initial configurations $u^+_k$, regardless to the degree of freedom contained in the initial configuration for the variable $u^-$. For this, the one-variable map for the small billiard, in its two symmetry-quotienting versions accounting for the epoch map and for the era-transition map, describes the physical trajectories, i.e. those implied by the evolution of the scale factors given by the solution to the Einstein field equations for the big billiard.\\
For this, the following definitions are obtained for the one-variable map of the quotiented small billiard, i.e. when all the permutations are considered
\begin{subequations}\label{smallprobper}
\begin{align}
&{\rm p}^{BKL}_{I 1-var}(k)\equiv {\rm p}^{BKL}_{I\ \ per}(k)\equiv \tfrac{{\rm p}^{BKL}_{I}(k)}{\langle {\rm p}^{BKL}_{I}(k)\rangle_{per}},\\
&{\rm p}^{BKL}_{II 1-var}(k)\equiv {\rm p}^{BKL}_{II\ \ per}(k)\equiv \tfrac{{\rm p}^{BKL}_{II}(k)}{\langle {\rm p}^{BKL}_{II}(k)\rangle_{per}},
\end{align}
\end{subequations}
where the denominators of (\ref{smallprobper}) are defined as
\begin{subequations}\label{smallprobdenomper}
\begin{align}
&\langle {\rm p}^{BKL}_{I}(k)\rangle_{per}\equiv\sum_{per(k)}{\rm p}^{BKL}_{I}(k)\\
&\langle {\rm p}^{BKL}_{I}(k)\rangle_{per}\equiv\sum_{per(k)}{\rm p}^{BKL}_{II}(k).
\end{align}
\end{subequations}
\\
\subsection{Probabilities for the unquotiented small billiard}
The unquotiented dynamics of the small billiard, i.e. a description of the small billiard dynamics where it is possible to take into account also the order of the Kasner transformation that allows to restore the description of periodic trajectories in the unquotiented big billiard, and then to generalize the reasoning for a comparison of the Kasner quotiented dynamics with the full unquotiented dynamics, where new features of the big billiard have been analyzed, i.e. the slight 'anisotropy' in the overall 'exploration' of the three corners of the big billiard according to the BKL statistics, which implies that short eras are most probable, and one-epoch ares the most favored.\\
For this, it is useful to generalize the definition of BKL probabilities for the small billiard, as far as the description of a trajectory of the unquotiented big billiard is concerned. As a result, by generalizing the definition of the previous sections, the following  probabilities are obtained
\begin{subequations}\label{smallprobsequencepk}
\begin{align}
&{\rm p}^{BKL}_{I}(p(n_1, n_2, ..., n_k))\equiv\int_{n_1-1+\tfrac{1}{n_2+\tfrac{...}{n_{pk}-1}}}^{n_1-1+\tfrac{1}{n_2+\tfrac{...}{n_{pk}}}}W^{I}( u^+)du^+,\\
&{\rm p}^{BKL}_{II}(n_1, n_2, ..., n_k)\equiv\int_{n_1-1+\tfrac{1}{n_2+\tfrac{...}{n_{pk}-1}}}^{\tfrac{1}{n_2+\tfrac{...}{n_{pk}}}}W^{II}(u^+)du^+,\\
\end{align}
\end{subequations}
where the integration extrema are given according to the definitions of the densities of measure $W^{\mu}(u^+)$ Eq.'s (\ref{densitymu}) specified for the small billiard.
%%%%%%%%%%%%%%%%%%%%%%%%%%%%%%%%%%%%%%%%%%%%
\subsection{The different statistical maps for the unquotiented small billiard}
The different statistical maps, i.e. the era-transitions maps, require a definition of BKL probabilities for the unquotiented small billiard dynamics, as far as its comparison with the unquotiented big billiard is concerned, for the specification of the role of the variable $u^+$ in the definition of trajecotires is implied. As a result, different trajectories can be compared, within the aim to describe the role of the variable $u^-$ in the description of the initial conditions for the asymptotic limit of the Einstein field equations towards the cosmological singularity, also according to the different order of the Kasner transformations needed to recast the equivalence between the quotiented big billiard and the unquotiented big billiard; this is possible by defining the following specifications for BKL probabilities for a repetition of $p$ times a sequence $k$,
\begin{subequations}\label{smallprobsymmpk}
\begin{align}
&{\rm p}^{BKL}_{I cycl}(pk)\equiv \tfrac{{\rm p}^{BKL}_{I}(pk)}{\langle {\rm p}^{BKL}_{I}(pk)\rangle_{symm}},\\
&{\rm p}^{BKL}_{II cycl}(pk)\equiv \tfrac{{\rm p}^{BKL}_{II}(pk)}{\langle {\rm p}^{BKL}_{II}(pk)\rangle_{symm}},
\end{align}
\end{subequations}
where the denominators of (\ref{smallprobsymmpk}) are defined according to a suitable symmetry operation on the elements of a sequence $k$ repeated $p$ times
\begin{subequations}\label{smallprobdenomsymmpk}
\begin{align}
&\langle {\rm p}^{BKL}_{I}(pk)\rangle_{symm}\equiv\sum_{symm(pk)}{\rm p}^{BKL}_{I}(pk)\\
&\langle {\rm p}^{BKL}_{II}(pk)\rangle_{symm}\equiv\sum_{symm(pk)}{\rm p}^{BKL}_{II}(pk).
\end{align}
\end{subequations} 
\subsection{The two-variable map in the unquotiented small billiard}
This way, different periodic trajectories of the unquotiented big billiard can be compared by considering them as at different stages between the iterations of the billiard maps on the same periodic trajectory, by the definitions 
\begin{subequations}
\begin{align}
&{\rm p}^{BKL}_{I \ \ 2-var}(pk)\equiv {\rm p}^{BKL}_{I \ \ cycl}(pk)\equiv \tfrac{{\rm p}^{BKL}_{I}(pk)}{\langle {\rm p}^{BKL}_{I}(pk)\rangle_{cycl}},\\
&{\rm p}^{BKL}_{II \ \ 2-var}(pk)\equiv{\rm p}^{BKL}_{II\ \ cycl}(pk)\equiv \tfrac{{\rm p}^{BKL}_{II}(pk)}{\langle {\rm p}^{BKL}_{II}(pk)\rangle_{cycl}},
\end{align}
\end{subequations}
with a suitable definition of the denominators, i.e.
\begin{subequations}\label{smallprobdenomcyclpk}
\begin{align}
&\langle {\rm p}^{BKL}_{I}(pk)\rangle_{cycl}\equiv\sum_{cycl(pk)}{\rm p}^{BKL}_{I}(pk)\\
&\langle {\rm p}^{BKL}_{II}(pk)\rangle_{cycl}\equiv\sum_{cycl(pk)}{\rm p}^{BKL}_{II}(pk).
\end{align}
\end{subequations}
\subsection{The one-variable map in the unquotiented small billiard}
Similarly to the previous analyses, the different singular trajectories of cosmological billiards can be compared according to the mechanisms which define the  transition between eras as a change of derivative for the non oscillating scale factor, according to the different patterns characterizing the evolution of the oscillating scale factors, by considering the classes of singular trajectories, which are described by different ordering of the eras containing a given sequence of epochs, and to compare them with the dynamics of the unquotiented big billiard.
\begin{subequations}\label{smallprobperpk}
\begin{align}
&{\rm p}^{BKL}_{I \ \ 1-var}(pk)\equiv {\rm p}^{BKL}_{I \ \ per}(pk)\equiv \tfrac{{\rm p}^{BKL}_{I}(pk)}{\langle {\rm p}^{BKL}_{I}(pk)\rangle_{per}},\\
&{\rm p}^{BKL}_{II \ \ 1-var}(pk)\equiv{\rm p}^{BKL}_{II\ \ per}(pk)\equiv \tfrac{{\rm p}^{BKL}_{II}(pk)}{\langle {\rm p}^{BKL}_{II}(pk)\rangle_{per}},
\end{align}
\end{subequations}
with the appropriate normalizing sums, i.e.
\begin{subequations}\label{smallprobdenomperpk}
\begin{align}
&\langle {\rm p}^{BKL}_{I}(pk)\rangle_{per}\equiv\sum_{per(pk)}{\rm p}^{BKL}_{I}(pk)\\
&\langle {\rm p}^{BKL}_{II}(pk)\rangle_{per}\equiv\sum_{per(pk)}{\rm p}^{BKL}_{II}(pk).
\end{align}
\end{subequations} 
\subsection{The two-variable map in the unquotiented small billiard}
The two-variable maps in the unquotiented small billiard read
\begin{subequations}\label{smallprobcyclpk}
\begin{align}
&{\rm p}^{BKL}_{I 2-var}(pk)\equiv {\rm p}^{BKL}_{I cycl}(pk)\equiv \tfrac{{\rm p}^{BKL}_{I}(pk)}{\langle {\rm p}^{BKL}_{I}(pk)\rangle_{cycl}},\\
&{\rm p}^{BKL}_{II 2-var}(pk)\equiv{\rm p}^{BKL}_{II cycl}(pk)\equiv \tfrac{{\rm p}^{BKL}_{II}(pk)}{\langle {\rm p}^{BKL}_{II}(pk)\rangle_{cycl}},
\end{align}
\end{subequations}
with a suitable definition of the denominators, i.e.
\begin{subequations}\label{smallprobdenomcycl}
\begin{align}
&\langle {\rm p}^{BKL}_{I}(pk)\rangle_{cycl}\equiv\sum_{cycl(pk)}{\rm p}^{BKL}_{I}(pk)\\
&\langle {\rm p}^{BKL}_{II}(pk)\rangle_{cycl}\equiv\sum_{cycl(pk)}{\rm p}^{BKL}_{II}(pk),
\end{align}
\end{subequations}
and where the dynamics regions of the restricted phase space $\mu$ are considered for the different number of Weyl reflection contained in the map for the quotiented small billiard, which is here compared to the unquotiented dynamics of the big billiard table as far as the iterations of the CB-LKSKS map are concerned.
%%%%%%%%%%%%%%%%%%%%%%%%%%%%%%%%%%%%%%%%
\section{A Stochastizing BKL dynamics for the small billiard\label{section9}}
%%%%%%%%%%%%%%%%%%%%%%%%%%%%%%%%%%%%%%%%%%%%%%%%%%
The most relevant step in the description of the stochastizing BKL dynamics of the small billiards consists in comparing the different asymptotic values of the BKL probabilities for the values of the variables $u^+$ and $u^-$, i.e. for the different dynamical subregions of the restricted phase space, where the small billiard map acquires a different number of reflections.\\
In the limit of a stochastic evolution of the dynamics, the probabilities for eras containing a given number of epochs become independent of the initial condition, and follow the asymptotic behavior analyzed for the comparison with the behaviors of the big billiard. For the definition of the order of the asymptotic behavior it is useful to analyze the trend assumed by these probabilities as a function of the number $n$ of epochs contained in a given era: the first relevant values are listed in Table \ref{tablebb}.\
%%%%%%%%%%%%%%%%%%%%%%%%%%%%%%%%%%\
\begin{table}
\begin{center}
    \begin{tabular}{ | l | l | l | l |}
    \hline
     $n$ & $P^{BKL}_\mu(n)$ & ${\rm p}^{BKL}_{I}(n)$ & ${\rm p}^{BKL}_{II}(n)$ \\ \hline 
     $1$ & $0.4150374989$ & $0.3390359531$ & $0.07600154597$ \\ \hline 
     $2$ & $0.1699250015$ & $0.08496250155$ & $0.08496250098$   \\ \hline 
     $3$ & $0.09310940485$ & $0.03227012477$ & $0.06083927943$   \\ \hline  
    \end{tabular}
\end{center}
\caption{\label{tablebb} Comparison of probabilities for the symmetry-quotienting mechanisms: BKL probabilities for the occurrence of eras containing $n$ epochs for cosmological billiards. The BKL probabilities $P^{BKL}$ refer to the big billiard, and are recalled in Eq. (\ref{probbb}), where the script $0$ is omitted. The BKL probabilities ${\rm p}^{BKL}_{\mu}(n)$ refer to the small billiard, and are defined in Eq.'s (\ref{smallprob}). The values of ${\rm p}^{BKL}_{I}(2)$ and ${\rm p}^{BKL}_{II}(2)$ are shown to be safely different for the reported accuracy of the softwer.}
\end{table}
%%%%%%%%%%%%%%%%%%%%%%%%%%%%%%%%%%%%%%%
As a result, it is convenient to define the stochastic limit of the BKL probabilities for the small billiard by a different approximation with respect to the procedure followed in the case of the big billiard; in particular, it is needed to specify the number $n$ of epochs contained in the era, as from the comparison of the numerical values listed in Table \ref{tablebb}, i.e.
\begin{subequations}\label{nonsaprei}
\begin{align}
&\lim_S {\rm p}^{BKL}_{I}(n\le 2)=\simeq \tfrac{1}{c_I}\tfrac{1}{n^2},\ \ \lim_S {\rm p}^{BKL}_{I}(n\ge 3)=\simeq \tfrac{1}{c'_I}\tfrac{1}{n^3},\\ 
&\lim_S {\rm p}^{BKL}_{II}(n\le 2)=\simeq \tfrac{1}{c_{II}}\tfrac{1}{n^3},\ \ \lim_S {\rm p}^{BKL}_{II}(n\ge 3)=\simeq \tfrac{1}{c'_{II}}\tfrac{1}{n^2}, 
\end{align}
\end{subequations}, 
where $c_I$, $c_{II}$, $c'{I}$ and $c'_{II}$ are the suitable coefficients defined in the series expansion of the BKL probabilities for the small billiard Eq.'s (\ref{smallprobexpansion}). Within all the steps of the stochastizing dynamics, these coefficients can be assumed to depend also on $n-j$, while this dependence should be completely removed within the framework of the full stochastized regime, which will be discussed in the next Section.\\
Differently from the big billiard dynamics, according to this comparison one is able to appreciate that the stochastization process of the dynamics of the small billiard is characterized not only by the independence of the probability for an era containing a given number of epochs to occur, but also by the number of epochs itself, which, on its turn, characterizes the different patterns of the evolution of the scale factors.
\subsection{Stochastizing BKL dynamics for the quotiented small billiard}
The most relevant steps in the stochastizing BKL dynamics for the small billiard in the symmetry-quotiented version of the statistical description a re illustrated by considering the choice (\ref{nonsaprei}) in the definitions of the limits for the different BKL probabilities
\be
\lim_S {\rm p}^{BKL}_{symm}(k)\equiv\tfrac{{\rm p}^{BKL}(k)}{n_{symm}(k){\rm p}^{BKL}(k)}
\ee
in which every term in the normalization is hypothesized to acquire the pertinent form.\\
\\
By considering each term that each term as consisting of the suitable composition of the choices of the probabilities (\ref{nonsaprei}) for each era independently of the initial conditions, to the extent as this hypothesis is permitted by (\ref{nonsaprei}), the stochastizing BKL dynamics is depicted as consisting of the processes described by the limits
\be
\lim_S {\rm p}^{BKL}_{symm}(k)\simeq\tfrac{{\rm p}^{BKL}(k)}{n_{symm}(k)\prod_{j=1}^{j=k}{\rm p}^{BKL}(k)}.
\ee
The asymptotic limit of each independent probability for the eras containing $n_j$ epochs leads to the most general expressions
\be
\lim_S P^{BKL}_{symm}(k)\simeq\tfrac{P^{BKL}(k)}{n_{symm}(k)c_{k\mu}\prod_{i=1}^{i=k}\tfrac{1}{n_i^2}},
\ee
where the coefficients $c_{k\mu}$ are assumed to have still a dependence both on $k$ and on the dynamical region of the restricted phase space $\mu$.
\subsection{Stochastizing BKL dynamics of the unquotiented small billiard}
After the analysis of the properties of the stochastization of the dynamics of the small billiard, it is possible to define the BKL probabilities for the small billiard according to a given symmetry operation on the generators of particular successions of eras.\\
As a result,
\be
lim_S {\rm p}^{BKL}_{\mu}(p k)\simeq p \prod_{i=1}^{i=k}{\rm p}^{BKL}_{\mu}(n_i).
\ee
\\
Similarly to the considerations followed for the stochastizing BKL dynamics in the quotiented case, BKL probabilities can be normalized according to a particular symmetry operation also for the repetition of $p$ times a sequence for the comparison with the full unquotiented dynamics of the big billiard, which is characterized by the determination of the order $p$ of the Kasner coefficient which allows for a complete closed geodesics.\\
For this, for the regions $\mu$, one defines these probabilities as
\be
\lim_S {\rm p}^{BKL}_{\mu symm}(pk )\equiv \tfrac{{\rm p}^{BKL}_{\mu}(p k )}{\langle {\rm p}^{BKL}_{\mu}{p k }\rangle_{symm}}.
\ee
where the denominators contain the sum
\be
\langle {\rm p}^{BKL}_{\mu}(p k )\rangle_{symm}=\sum_{p symm{k}}{\rm p}^{BKL}_{\mu}(k ).
\ee
and where the suitable dependence of the coefficients of the implied Taylor expansion is still allowed to depend on $p$, $k$ and $\mu$, as from the correct generalization of (\ref{nonsaprei}), i.e. by posing $c^\mu_{p,k}$. It is not specified on purpose, here and in the previous analyses, whether the explicit dependence of these coefficients is represented by $c^\mu_{p\{\ k\}\ }$ or $c^\mu_{\{\ pk\}\ }$, according to the different dependence this factor, form (\ref{nonsaprei}) can acquire during the (still unexplored) steps of the stochastizing BKL dynamics before the full stochastized regime is completely established.
\section{Stochastized BKL dynamics of the small billiard\label{section10}}
The stochastic limit $S$ for a BKL probability of the small billiard under the billiard map (\ref{tsb}) is defined as
\be\label{nonsaprei1}
\lim_S {\rm p}^{BKL}_{\mu}(k)\simeq\prod_{i=1}^{i=k}{\rm p}^{S}_{\mu}(n_i)
\ee
for the different subregions $\mu$ of the restricted phase space, where the stochastic limit of the BKL probabilites for the small billiard is supposed to hold at each iteration of the billiard maps, independently of the evolution of the system, as
\be\label{nonsaprei2}
\lim_S {\rm p}^{BKL}_{\mu}(n)\equiv {\rm p}^{S}_{\mu}(n)
\ee
where the presence of only one epoch $n$ recalls this property.
\subsection{The stochastized dynamics for the BKL probabilities of the quotiented small billiard}
The application of the stochastic properties acquired by the BKL dynamics after the definitions of (\ref{nonsaprei1}) and (\ref{nonsaprei2}) to the stochastic limit for a BKL probability normalized according to a symmetry operation on the generators that compose the iterations of the big billiard map in the case of the small billiard map (\ref{tsb1}), according to the definitions (\ref{smallprobdenomsymm}) of the denominators that normalize (\ref{smallprobsymm}) is obtained 
\be
\lim_S\langle {\rm p}^{BKL}_{\mu}(k)\rangle_{symm} \simeq n_{symm}(k)\lim_S{\rm p}^{BKL}_{\mu}(k),
\ee
where the following limit for a stochastization of the dynamics is posed
\be
\lim_S\langle {\rm p}^{BKL}_{\mu}(k)\rangle_{symm} \simeq c_{\mu}^N n_{symm}(k)\prod_{j=1}^{j=k}\tfrac{1}{n_j^N}
\ee
with the general degeneracy prefactor $n_{symm}(k)$, and no dependence of the $c_\mu^N$ on $p$ and $k$ is still allowed, but the overall dependence of the factor on $N$, with $N=2$ or $N=3$, according to the asymptotic expansion of the probabilties (\ref{nonsaprei}) with respect to the content of epochs in each era.
\subsection{The stochastized dynamics for the BKL probabilities of the unquotiented small billiard}
The limit to a stochastic process for the unquotiented dynamics of the small billiard is characterized by the assumption
\be
\lim_S{\rm p}^{BKL}_{\mu symm}(p\{\ k\}\ )\simeq pc_{\mu}^N\prod_{j=1}^{j=k}\tfrac{1}{n_j^N},
\ee
where the normalization with respect to a symmetry operation is therefore
\be
\lim_S\langle {\rm p}^{BKL}_{\mu}(p\{\ k\}\ )\rangle_{symm}=n_{symm}(pk)c_{\mu}^N\prod_{j=1}^{j=k}\tfrac{1}{n_j^N},
\ee
i.e. the denominators are characterized by the presence of the degeneracy prefactor $n_{symm}(pk)$ which depends on the order of the Kasner transformation $p$.\\ 
%%%%%%%%%%%%%%%%%%%%%%%%%%%%%%%%%%%%%%%%%%%%%%%%%%%%%%%
As a result, the implementation of this description seems differently easier to handle, and can allow for a discussion of the asymptotic (in the limit to a stochastization of the dynamics for the) behavior of the small billiard, in comparison with the description of the small billiard map implemented for the variable $u^+$ and $u^-$ on the restricted phase space, where eras and epoch for the small billiard correspond to  subdomains delimited by curvilinear functions, all the specifications needed for the region $I$ and all the correction proposed for the region $II$ allow one to continue to define probabilities for squared boxes corresponding still to the number $n$ of epochs in each era in the restricted phase space, according to Figure \ref{figura2}.
\section{Comparisons of the billiard maps\label{section11}}
It is now possible to analyzed the different description of the dynamics of the big billiard, that of the small billiard, the implementation of the one-variable maps and the implementation of the two-variable map, the Kasner quotiented dynamics and the full unquotiented dynamics, with the understanding that the iterations of the BKL map leads to a stochastization of the dynamics.
As a result, the information encoded in the small billiard maps as far as the evolution of the scale factors is concerned can be transferred to the description of the unquotiented big billiard, both in the early time regime, for a stochastizing dynamics and for the lull stochastic limit. One learns that the different epochs are described by a probability distribution that takes into account the number of Weyl reflections contained in the billiard maps on the UPHP, and this probability distributions are specified by the number $n$ of epochs contained in the different eras.\\
In particular, according to the series expansion of the BKL probabilities for the small billiard, a factorization can be stated for the probabilities of sequences of eras for the big billiard, according to the BKL probabilities for the small billiard
\be\label{order1}
P^{BKL}(k)\simeq\left(\prod_{i=1}^{i=2}{\rm p}^{BKL}_{I}(n_i)\right)+\left(\prod_{j=2}^{j=k}{\rm p}^{BKL}_{II}(n_j)\right)
\ee
where the equivalence holds for all the orders $\mathcal{O}\left(\tfrac{1}{n^i}\right)$, as far as the Taylor series expansion of $P^{BKL}(n)$ is concerned.\\
\\
When the full unquotiented dynamics of the big billiard is considered, for a repetition of $p$ times a sequence $k$ of eras, the following factorization holds
\be\label{order2}
P^{BKL}(pk)\simeq\left(\prod_{i=1}^{i=2}{\rm p}^{BKL}_{I}(n_i)\right)+\left(\prod_{j=2}^{j=pk}{\rm p}^{BKL}_{II}(n_j)\right),
\ee
which holds for all the orders $\mathcal{O}\left(\tfrac{1}{n^i}\right)$.\\
\\
As a consequence of (\ref{order1}) and (\ref{order2}), the equivalence between the big billiard and the small billiard is therefore proven not only for the exact BKL dynamics, but also as far as the stochastized BKL dynamic sis concerned.\\
\\
When a given symmetry operation on the generators of the reflections is taken into account, the BKL probabilities specify according to the results reported in Table \ref{tablebb}, illustrated also in Figure \ref{figura3}, for which ${\rm p}^{BKL}_{II}(n)>{\rm p}^{BKL}_{I}(n)$ for $n=1, 2$, while ${\rm p}^{BKL}_{II}(n)<{\rm p}^{BKL}_{I}(n)$, where the strict inequalities assure the validity of the following limiting expressions for the factorization of the BKL probabilities for the big billiard with respect to the stochastic limit of those of the small billiard, as far as the different versions of the statistical BKL maps are concerned:
\be
P^{BKL}_{symm}\equiv\tfrac{\left(\prod_{n_i=1,2, i=1, ..., k}{\rm p}^{BKL}_{II}(n_i)\right)\left(\prod_{n_j>2, j=1, ..., k}{\rm p}^{BKL}_{I}(n_j)\right)}{\langle \left(\prod_{n_i=1,2, i=1, ..., k}{\rm p}^{BKL}_{II}(n_i)\right)\left(\prod_{n_j>2, j=1, ..., k}{\rm p}^{BKL}_{I}(n_j)\right)\rangle_{symm}},
\ee 
When the repetition of $p$ times a trajectory is taken into account, in the unquotiented dynamics the previous results most generally specify as 
\be
P^{BKL}_{symm}(pk)\equiv\tfrac{\left(\prod_{n_i=1,2, i=1, ..., pk}{\rm p}{BKL}_{II}(n_i)\right)\left(\prod_{n_j>2, j=1, ..., pk}{\rm p}^{BKL}_{I}(n_j)\right)}{\langle \left(\prod_{n_i=1,2, i=1, ..., pk}{\rm p}^{BKL}_{II}(n_i)\right)\left(\prod_{n_j>2, j=1, ..., pk}{\rm p}^{BKL}_{I}(n_j)\right)\rangle_{symm}}.
\ee 
According to these most general expressions, any degree of stochasticity for the evolution of the BKL dynamics towards the asymptotic limit of the sully stochastized regime can be taken into account.
By means of these considerations, it is straightforward to extend the definition of BKL probability for an era to contain a number $n$ of epoch by encoding in it the properties of the evolution of the components of the metric tensor as defined by the number of Weyl reflections on the UPHP, let's say $n_W$, for the  probabilities, as it directly follows from (\ref{order1}) and (\ref{order2}), which establish the equivalence between the big billiard and the small billiard. In particular, given
\be
P(n, n_W=1)\equiv \tfrac{{\rm p}^{BKL}_{I}(n)}{P^{BKL}(n)}
\ee
and
\be
P(n, n_W=2)\equiv \tfrac{{\rm p}^{BKL}_{II}(n)}{P^{BKL}(n)}
\ee
the new BKL probability for the big billiard nicely rewrites as
\be
P^{BKL}(n, n_W)\equiv P^{BKL}(n)\left(P(n, n_W=1)+P(n, n_W=2)\right)
\ee
according to the versions of the small billiard map (\ref{tsb}) on the UPHP.
%%%%%%%%%%%%%%%%%%%%%%%%%%%%%%%%%%%%%%
%%%%%%%%%%%%%%%%%%%%%%%%%%%%%%%%%%%%%%%%%%%%
\section{Singular trajectories\label{section12}}
The analyses developer in the previous Sections allows one to gain insight about the nature of singular trajectories.\\
The importance of the classification of singular trajectories in the dynamics of cosmological billiards, as far as the original BKL dynamics is concerned, as well as in the comparison of the previous description with the new structures found in the higher-dimensional descriptions, for which the usual BKL limit is recovered in $4=3+1$ dimensions, and for which the description of the more general inhomogeneous case is obtained by the definition of the dominant symmetry walls, has been recently recalled in \cite{bel2009}.\\
By means of the insight gained throughout this investigations, it is possible to infer that singular trajectories can find a characterization by the implementation of the one-variable map, i.e. by considering the probability associated to the occurrence of a given finite sequence, and then by comparing the probability of this sequence with those described by the same number of epochs in a different re-arrangement of th e sequence of epochs. this way, it is possible to compare the features of different singular trajecotires within a stochastized dynamics. The results obtained for the definition of the stochastization of the dynamics for the probabilities associated with the one-variable map therefore describe the different probabilities for which singular trajectories are found.\\
\\
Furthermore, the analysis of singular trajectories can be followed also by taking into account the different umber of Weyl reflections contained in the BKL maps for the variable $z$ on the UPHP. As a result, it is possible to rewrite the corresponding BKL probabilities as a product of the corresponding probabilities of the small billiard, where the extents to which their Taylor series is valid has to be taken into account.
%%%%%%%%%%%%%%%%%%%
\section{Anisotropic Sky Patterns\label{section13}}
In \cite{barrowlevin}, the possible microwave patterns observable are demonstrated to be connected with the group-theoretical properties of the three-geometries that characterize the spatial part of the metric tensor, and special attention is paid to possible applications to the Bianchi classification. In particular, following \cite{hadamard}, the chaotic properties of the geodesic flow on compact surfaces of negative curvature were clarified in \cite{barrowlevin} for the cosmological setting of small universe, and adapted the description of such system for the purpose of the detection in the CMB spectrum of the properties of such a geodesic flow for homogeneous cosmologies and developed a paradigm for a possible application of the same analysis for some Bianchi cosmologies with small anisotropies, according to the schemes outlined in \cite{novikov68} \cite{hawking69} \cite{collins73} \cite{doro75} \cite{barrow85} \cite{barrow851}. For this, the precise group theoretical structure of the three-geometry of the metric tensor is needed.\\
\\
In the present work, the geodesic flow of the most general Bianchi universe has been analyzed after the solution of the Hamiltonian constraint for the Einstein field equations, such that there is one degree of freedom less. Within this description, not only trajectories have been classified, but suitable expressions for  probabilities for the occurrence of these geodesics has been provided, according to the extent of stochasticity in which the dynamics has to be analyzed. The degree of stochasticity of the system is needed to be known, when a suitable quasi-isotropization mechanism has to be considered, such that the actual little value for the observed anisotropy of the universe is theoretically recast. For this, a precise description of the probability for a particular class of geodesics is possible after the determination of the regime (i.e. pure BKL regime, stochastizing BKL regime, or stochastized regime) at which the quasi-isotropization mechanism is hypothesized to happen. Indeed, in the description of \cite{barrowlevin}, the only assumption made for the work is that of a small universe: in the vicinity of the cosmological singularity, a more precise determination of the stochasticity reached by the universe should be clarified by the comparison with present observations.\\
\subsection{Anisotropy and Stochasticity}
Nonetheless, it is possible to transfer the information gained after the implementation of the Hamiltonian constraint to classes of geodesics in the description of the full chaotic motion, where a three-geometry in the target space of the billiard ball is considered, for the complete characterization of the properties of the spatial part of the metric tensor. In fact, the properties of the motion of the billiard ball in the unprojected target space exhibits extremely complicated features, such that several features outlined in the analysis of the UPHP are lost.\\
For this, the consideration of classes of geodesics instead of single geodesics can be helped by the consideration of classes of BKL probabilities in the stochastizing BKL paradigm and in the stochastized regime, which, on its turn, is feasible after the knowledge of the stochasticity acquired by the system in the vicinity of the cosmological singularity before the application of the quasi-isotropization mechanisms.\\
From a different point of view, the present classification is effective in characterizing the singular trajectories, periodic trajectories and irrational trajectories following \cite{hadamard} as in \cite{barrowlevin} at each extent of stochasticity, such that more information can be gathered for the description of classes of geodesics in the unprojected model.\\
Transferring this information to the unprojected model has to be compared with choosing a suitable Poincar\'e surface of section for the determination of the Poincar\'e return map for the geodesics. As studied in \cite{lecian}, the properties of the Poincar\'e return map depend on this choice, and the determination of the BKL trajectory $u^+$ is relevant in this process. The determination of the initial conditions $u^+$ and $u^-$ for small universe is a feature which holds at the classical regime, in the semiclassical limit and in the quantum version of the model, specifying, for the determination of the Poincar\'e return map, the extent at which possible quantum features of the gravitational interaction modify the classical regime can be directly added to the determination of the effects of the quasi-isotropization mechanism acting on a certain degree of stochastization of the dynamics.\\
\\ 
The presence of epochs joining different sides of the billiard table in the desymmetrized version of the dynamics is due to the fact that the Poincar\'e return map for the small billiard is defined in a surface of section different from one defined by an entire gravitational wall of the big billiard table. This way, epoch joining different sides of the billiard are equiparated after mapping them on the regions of the phase space, which correspond to epochs of the same type; this mapping implies the presence of an extra reflection, corresponding to that contained in the pertinent Kasner transformation. This further specification of the BKL probabilities can therefore better characterize the description of the present universe from the experimental point of view. In particular, this specification is essential in the description of the homogeneity of the observed universe, as the description involving half the gravitation wall and the symmetry walls corresponds to the asymptotic limit of the most general inhomogeneous models, while the presence of the three gravitational walls, for which an entire wall of the billiard table can be chose as a surface of section, corresponds to the pure gravitational case.\\
\subsection{A different characterization of the chaotic properties} A different characterization of chaotic systems can be achieved by the definitions of the entropy. The metric entropy \cite{kolmogorov1}, \cite{kolmogorov2}, \cite{sinaient} and the topological entropy \cite{topologicalentropy} are used for the characterization of chaotic systems, whose statistical properties can be analyzed without any explicit dependence of the properties of the system on time. The scale factors which constitute the Kasner solution, which are functions of time, define the free-flight geodesics evolution of the billiard ball; as energy is conserved within the dynamics, the geodesics evolution of the billiard trajectories can be considered as punctuated by each bounce, which corresponds to a change in the dynamics of the Kasner exponents. Accordingly, the billiard ball is described as moving at constant velocity; nevertheless, the number of epochs contained in each era define the different angular velocities at which the billiard ball bounces on the boundaries of the billiard table, and has a precise physical interpretation at the classical level, at the quantum regime and in the semiclassical limit. Within this framework, in \cite{barrowpr}, the Kolmogorov-Sinai entropy and the topological entropy have been evaluated for cosmological billiards\footnote{A parammetrizzation of the endpoints has been proposed in \cite{Kirillov:1996rd}, such that the resulting description burdens the complete statistical description of \cite{Chernoff:1983zz} with an explicit dependence on the time, for which a specification of metric entropy or topological entropy is not possible, and for which the stochastization process of the dynamics is not treatable, such that the relevance of initial conditions expressed in $u^-$ is undermined.}.
\subsection{Small anisotropies}
A classification of the trajectory is possible not only according to the initial conditions for the Einstein field equations, but also according to the definition of BKL probabilities for the different implications of these initial conditions, and also for the degree of stochasticity reached by the system on the UPHP.\\
The unprojected model is demonstrated to exhibit extremely more complicated features, for which a definition of periodicity is different with respect to that fixed for the UPHP.\\
A determination of the age of the universe at which the quasi-isotropization mechanism has started played a role observable detectable in the present observations \cite{9a},\cite{9b}, \cite{9c} implies the degree of stochasticity according to which the classification of the BKL probabilities has to be considered.\\
The properties of the oriented endpoints in determining the properties of the motion are independent to whether the quasi-isotropization mechanism has started playing its role at the quantum regime, in the semiclassical limit or in the classical regime. The specification of a particular statistical map, i.e. the choice of the characterization of a class of trajectories according to the variable $u^-$ or the choice to marginalize it therefore depends on the stochastic properties acquired by the system after the iterations of the billiard map, for which its is possible to discriminate about the importance of the 'memory' of the systems about the 'past' evolution encoded in the variable $u^-$.\\
The combination of all these possibilities depends on the assumption made about the evolution of the present universe, and a detection of the features of the anisotropy is fundamental within this framework. The implications, both from  theoretical point of view, and as far as possilble observations a reconcerned, are discussed in some details in \cite{anis1}, \cite{anis2}, \cite{anis3}, \cite{anis4}, \cite{anis5}, \cite{anis6}, \cite{anis7}.\\
\\
On the other hand, the definition of these entropies is complementary to the present investigation, in the sense that the topological entropy is based on the qualification of different trajectories according to how different two trajecotires become after a suitable number of iterations of the billiard, where the number of iterations of the billiard map defines the age of the universe if combined with the number of epochs contained in each era, whose length corresponds to a suitable time duration, such that the definition of the metric entropy for these systems corresponds to the parametrization of a given age of the universe through the length of the Kasner eras as specified by the time evolution of the scale factors.\\
\\
Combining these notions allows one to define the extant of (in)dependence of the trajectories on their initial condition, at a given degree of stochasticity of the dynamics, which also defines the independence of the 'past' evolution contained in the variable $u^-$. The choice of this kind of specifications allows one to clarify the age of the universe at which a quasi-isotropization mechanism has happened, through the characterization of the resulting model by means of the values of anisotropies observed.\\
For this, it is possible, in principle, to specify if the quasi-isotropization mechanism has taken place during the quantum regime, and can therefore be ascribed to possible quantum features of the gravitational interaction below the Planck scale, or during the semiclassical regime, for which these possible quantum features have started acquiring classical properties at the Planck scale, or already during the classicalized regime, such that the quasi-isotropization mechanism can be identified with other, possibly 'external' or 'non-gravitational', causes. All these hypothesis have also to be matched with the possibility of any isotropization mechanism arising from a compactification of the higher number of dimensions, where new structures have been found, which could have modified the anisotropic content of the metric tensor according to the age of the universe, at which the compactification mechanism has started exerting its influence \cite{scalvec}.\\
For this, it is crucial to recall that the analysis of the Einstein field equations alone is not sufficient for these kinds of discriminations, as all this effects, within the framework of cosmological billiards, are mapped to the UPHP via a modification ascribed to any kind of term present on the rhs of the Einstein field equations. Nevertheless, the exact age of the universe at which the quasi-isotropization mechanism starts playing a role as strong as to modify the BKL original dynamics can be traced in the full unprojected model by looking for the corresponding degree of stochastization of the BKL dynamics. The description of the stochastizing BKL dynamics, as well as its fully stochastized limit, assume therefore a very relevant role within this analysis.
%%%%%%%%%%%%%%%%%%%%%%%%%%%%%%%%%%%%%%%%%%%%%%%%%%%%%%%
%%%%%%%%%%%%%%%%%%%%%%%%%%%%%%%%%%%%%%%%%%%%%%% 
\section{Concluding remarks\label{section14}}
The classification of the initial conditions for these systems corresponds to the classification of the values acquired by the variable $u^+$ analyzed according to its continued-fraction decomposition.\\
\\
For periodic trajectories, the normalized density of invariant measure has been defined in the different symmetry-quotienting mechanisms.\\
According to the features of BKL probabilities, different probabilities have been defined, able to encode the information about periodic initial conditions in the Selberg trace formula, according to the different aspects of the dynamics.\\
\\
The BKL statistics can be implemented for cosmological billiards as a one-variable map and as a two-variable map. The second case is obtained form the first case by marginalizing one variable. The variable that is marginalized is interpreted, in the case of the two-variable map, as the variable which encodes the past evolution of the billiard, and its statistical relevance is due to the possibility to find a precise statistical characterization of the billiards map for the case of the two-variable map, and a precise statistical characterization of their stationary limit. When this variable is marginalized form the statistical map, it is usually attributed the meaning of an extra degree of freedom within the framework of the specification of the initial conditions. Within the present analysis, the elements of the conjugacy subclasses of the composition of matrices which result as periodic trajectories with the same hyperbolic length have been related to the evolution of the solution to the Einstein field equations, such that it has been possible to establish that the cyclic permutations of the generators that compose a periodic sequence are consistent with the two-variable BKL maps, as the cyclic permutations allow one to compare different periodic orbits after the iterations of the billiard maps. On the contrary, exchange permutations among the elements of the composition of matrices would correspond to a reshuffling of the periodic sequences, which would imply a modification for the past evolution of the billiard system: for this reason, specific probabilities have been defined, which allow for a comparison of periodic orbits under the iterations of the billiard map only. Differently, the exchange permutations among the generators of the composition of matrices which allow one to reshuffle a periodic sequence are comprehended within the same definition of probabilities for the one-variable maps, and are therefore summed within the same BKL probability.\\
\\
The sum over the periodic initial conditions has therefore been reconducted to a sum over the periodic sequences, which are accounted for the elements of the conjugacy subclasses of the composition of matrices obtained for the periodic initial conditions. As a result, different initial conditions originate different terms in the sum over the energy levels, such that, differently from the analysis of hyperbolic billiards within the framework of the mathematical characterization of the properties of the group that generate the symmetries of billiard systems, the elements of the conjugacy subclasses of the groups that describe the dynamics of cosmological billiards are weighted according to different procedures. As a result, the influence of the initial conditions in the characterization of particular sets  
 of trajectories is clearly outlined. In the sum over the symmetry operations on the matrices that generate the periodic sequences, the statistical BKL maps are demonstrated to be able to outline different features of cosmological billiards. The presence of different subregions of the restricted phase space, where the statistical maps imply different transformation for the statistical variables, the different effects of the comparison of the unquotiented dynamics with the different symmetry-quotienting mechanisms, the relevance of the different initial conditions in the determinations of non-trivial prefactors for the same values of the hyperbolic lengths of the closed trajectories, are all features which have been clearly outlined. In particular, this characterization of the dynamics allows one to infer that different initial values for the statistical variables are able to produce different limits in the stochastization of the dynamics.\\
 As a result, the probabilities here defined acquire a higher value for those trajectories, whose statistical BKL probability is the highest. These trajectories are the lowest-order periodic geodesics, i.e. those characterized by a small ($n\sim1$) number of epochs in each era. The number of epochs in each era corresponds to the values acquired by the Kasner coefficients during the evolution of the dynamics. For this, one directly infer that the evolution of those models, whose behavior in the asymptotic limit to the cosmological singularity is one obtained as the same asymptotic limit of the Bianchi IX model, is characterized by these most probable periodic orbits. According to the Ljyapunov analysis of the dynamics, the relevance of periodic orbits is such that a generic geodesics can be approximated, within the range of the perturbations to the values of the statistical maps, to a periodic one.\\
Moreover, it is now possible to extend the original BKL statistic by adding the notion of Weyl reflection to each definition of probability for the content of epochs in each era, where this notion directly encodes the most relevant features of the evolution of the scale factors.\\  
\\
The paper has been organized as follows.\\
In Section \ref{section2}, the features of cosmological billiards which are relevant in the present investigation have been introduced.\\
In Section \ref{section3}, the trajectories of cosmological billiards have been classified, within the schematization of the solution of the Einstein field equations in the asymptotic limit towards the cosmological singularity, are classified according to the initial conditions of the Einstein field equations themselves, which admit a straightforward parametrization in the discretized statistical schematization of the corresponding dynamics.\\
In Section \ref{section4}, new invariant densities of measure have been defined for the small billiard.\\
In Section \ref{section5}, new probabilities for the big billiard have been defined, which are normalized according to the symmetry operations that the definitions of the one-variable BKL map and the two-variable BKL map describe, and a physical interpretation is given for these new definitions.\\
In section \ref{section6}, the stochastizing process of the dynamics of the big billiard is discussed according to the different statistical maps and the corresponding new probabilities.\\
In Section \ref{section7}, the full stochastic regime of the big billiard is described.\\
In Section \ref{section8}, probabilities for the small billiard are defined according to the new densities of invariant measures defined in the previous Section \ref{section4}, as well as the new BKL probabilities normalized according to the one-variable maps ans to the two-variable map.\\
In Section \ref{section9}, the stochastization of the dynamics of the small billiard has been analyzed, and the difference with the features of the the very same stochastizing dynamics of the big billiard have boon outlined.\\
In Section \ref{section10}, the full stochastic regime of the small-billiard dynamics has been described, and the difference with the same limit of the big-billiard have been clarified.\\
In Section \ref{section11}, singular trajectories have been commented within the previous results.\\
In Section \ref{section12}, within the framework of the equivalence of the big billiard with the small billiard, the dynamics of the big billiard has been described by means of the BKL probabilities defined for the small billiard, and the BKL statistical analysis about the most probable content of epochs for each BKL era has been enriched by the notion of Weyl reflections, which characterize the small billiard map on the UPHP.\\
In Section \ref{section13}, the connection of this analysis with the actual observed values of the CMB have been proposed..\\
Brief concluding remarks end the paper.
%%%%%%%%%%%%%%%%%%%%%%%%%%%%%%%%%%%%%%%%%%%%
%%%%%%%%%%%%%%%%%%%%%%%%%%%%%%%%%%%%%%%
%%%%%%%%%%%%%%%%%%%%%%%%%%%%%%%%%%%%%%%%%%%%%%%%%%%
\section*{Acknowledgments}
The work of OML was partially supported by the research grant 'Reflections on the Hyperbolic Plane' from the Albert Einstein Institute - MPI, Potsdam-Golm, and partially by the research grant 'Classical and Quantum Physics of the Primordial Universe' from Sapienza University of Rome- Physics Department, Rome.\\
OML acknowledges the Albert Einstein Institute- MPI warmest hospitality during the corresponding stages of this work.\\
OML is grateful to Prof. H. Nicolai for encouraging the investigation of periodic phenomena in cosmological billiards, and to Prof. P. De Bernardis for suggesting to analyze the chaotic features of the asymptotic Bianchi IX solutions.\\
OML kindly thanks A. Kleinschmidt for a strictest discussion of these results.\\
The environment Maple was used.
%%%%%%%%%%%%%%%%%%%%%%%%%%%%%%%%%%%%%%%%%%%%%%%

\end{document}